\documentclass[12pt]{iopart}

\usepackage{booktabs}
\usepackage{multirow}
\usepackage{graphicx}
\usepackage{ulem}
\usepackage{hyperref}
\usepackage[T1]{fontenc}

\begin{document}

\title{Reaching sub-millisecond accuracy in stellar occultations and artificial satellites tracking}

\author{K. Kamiński$^1$, C. Weber$^2$, A. Marciniak$^1$, M. Żołnowski$^3$ \& M. Gędek$^3$}

\address{$^1$Astronomical Observatory Institute, Faculty of Physics, A. Mickiewicz University, Słoneczna 36, 60-286 Poznań, Poland.}
\ead{chrisk@amu.edu.pl}
\vspace{10pt}
\address{$^2$IOTA/ES e.V., Am Brombeerhag 13, D-30459 Hannover, Germany.}
\ead{camera@iota-es.de}
\vspace{10pt}
\address{$^3$6 Remote Observatories for Asteroid and Debris Seaching, Godebskiego 55A, 31-999, Kraków, Poland.}
\ead{michal.zolnowski@6roads.com.pl }

\vspace{10pt}
\begin{indented}
\item[]January 2023
\end{indented}

\begin{abstract}
   {In recent years there appeared a need for astronomical observations timed with sub-millisecond accuracy.
    These include e.g. timing stellar occultations by small, sub-km or fast Near Earth Asteroids, but also
    tracking artificial satellites at Low Earth Orbit using optical sensors.
   }
   {Precise astrometry of fast-moving satellites, and accurate timing of stellar occultations have parallel needs,
    requiring reliable time source and good knowledge of camera delays. Thus a need for an external device that would enable
    equipment and camera testing, to check if they reach the required accuracy in time. }
   {We designed, constructed and thoroughly tested a New EXposure Timing Analyser (NEXTA): a GNSS-based precise timer (Global Navigation Satellite System),
   allowing to reach the accuracy of 0.1 millisecond, which is an order of magnitude better than in previously available tools.
   The device is a simple strip of blinking diodes, to be imaged with a camera under test and compare imaged time with
   internal camera time stamp.}
   {Our tests spanned a range of scientific cameras widely used for stellar occultations and ground-based satellite tracking.
    The results revealed high reliability of both NEXTA and most of the tested cameras, but also pointed that practically all cameras had
    internal time bias of various level.}
   {NEXTA can serve the community, being easily reproducible with inexpensive components.
    We provide all the necessary schemes and usage instructions.}
\end{abstract}


\maketitle

\section{Introduction}

 Certain observations in astronomy require sub-millisecond timing precision, either due to 
 short duration of the studied phenomena, or due to rapidly changing position of the studied bodies. 
 An example for the first are stellar occultations by small asteroids, and for the latter, the artificial satellite tracking 
 by optical, ground-based sensors. In the era of Gaia mission catalogues \cite{Gaia_EDR3} the predictions of stellar occultations 
 has recently faced unprecedented improvement, enabling to register occultations by only km-sized bodies 
 or even by Near-Earth Asteroids (NEAs), as exampled by recent successful occultation campaigns on Apophis and Phaethon, 
 being around two hundred meters and two kilometers in diameter, respectively \cite{Dunham2020}. Phaethon is 
 the target of JAXA's DESTINY+ mission, so knowing as much as possible about the target properties prior to the mission 
 is a key issue. Basing on recently acquired possibilities in occultation studies, an ACROSS 
 project has been launched (Asteroid Collaborative Research via Occultation Systematic Survey\footnote[1]{\url{https://lagrange.oca.eu/fr/home-across}}) aiming to regularly observe stellar occultations by small NEAs, 
 including binary Didymos -- target of the DART (NASA) and Hera (ESA) missions. 

 Stellar occultations by asteroids are among the most accurate methods to determine asteroid sizes. 
 The technique is quite simple, yet powerful: one only needs to precisely measure the moments of disappearance and reappearance 
 of a star occulted by a minor body from the solar system. The yield of occultation observations is greatest when 
 done from a network of observers optimally positioned within the predicted shadow path.
 Despite its simplicity, the ``resolving power'' of this method lies between the possibilities of a space telescope and in-situ studies 
 by a dedicated space mission. For example the diameters of Ceres and Vesta determined 
 from multi-chord stellar occultations observed with small telescopes from the ground are within 1\% from the direct measurements 
 made by Dawn spacecraft \cite{Herald2020}. 
 The most widespread technique to determine asteroid sizes however uses their absolute H magnitude and assumed albedo \cite{Harris1997}. 
 As such it can be off by even 50\%. More precise method, to use asteroid infrared fluxes from space observatories 
 can reach 30\% uncertainty (or the discrepancy between various missions results), being typically of the order of 10\% -- 20\%
 for Simple Thermal Model, with unknown spin and assumed spherical shape \cite{Harris2002}, 
 and decreasing to 5\% or less, when using Thermophysical Modelling with detailed spin and shape models \cite{Delbo2015}. 

 Occultation events also improve the astrometry, of both involved bodies with even milliarcsecond accuracy \cite{rommel_stellar_2020}, 
 enabling e.g. the Yarkovsky drift measurements \cite{Dziadura2022}. They also facilitated the discovery and studies on seasonal changes in the atmospheric profiles of distant dwarf planet Pluto \cite{Hubbard1988, Arimatsu2020}. 
 Such events also enable to discover rings \cite{Ortiz2017} and natural satellites of minor bodies, as recently exampled by the 
 first confirmed detection of a moon orbiting minor planet (4337) Arecibo, in two independent occultation events \cite{Gault2022}, 
 later also confirmed by photocenter offset of this body detected in Gaia mission data \cite{Tanga2022}.
 Typically, stellar occulations by asteroids are capable of breaking the inherent symmetry of two mirror pole solutions 
 from lightcurve inversion, and confirming the shape features of their 3-D shape models, or pointing to shape model areas that 
 still need some improvement \cite{Durech2011}. For large, Trans-Neptunian Objects, occultations amended with other observations  
 can lead to density estimates \cite{Ortiz2020}, allowing compositional studies of bodies so distant that often 
 invisible to small telescopes (it is only required to see the occulted star). Last but not least, stellar occultations 
 by solar system objects regularly unravel binary nature of stars, and give insight into brightnesses and separation of the stellar components 
 \cite{Herald2020}.

In recent years a rapid growth in the number of artificial Earth satellites is observed, reaching over 25000 in March 2022 in Space Track catalogue \cite{Cowardin2022}. Among them the population of Low Earth Orbit (LEO), with altitudes above the ground below 2000 km, is the most numerous, comprising about 60\% of all catalogued objects. Although typically LEOs are mostly tracked using ground based radars, there is a growing need for using optical telescopes for tracking and survey observations of them as well. This is partly because of the necessity of monitoring orbital perturbations \cite{Kaminski2018} and partly because of growing interference of satellite streaks with astronomical observations \cite{Michalowski2021}. Both areas would benefit from the increase of the accuracy of satellite position which can be obtained by combining range and Doppler measurements from radars and laser ranging stations with position measurements from optical telescopes. In order to reach the accuracy of astrometric measurements of LEOs at the level of an arcsecond it is necessary to assure an accurate image timing down to sub-millisecond level. This is because these satellites are typically observed at angular velocities of hundreds and thousands of arcsec/s. Optical observations of higher satellites usually include calibration targets, with accurately known orbits, such as navigational satellites. By comparing the predicted and observed positions a time-bias (the difference between the recorded image timing and the actual image timing) is derived. It is not uncommon for an astronomical camera to have a time-bias of the order of tens or even hundreds of milliseconds and sometimes the time-bias can change on a daily basis. Unfortunately, navigational satellites have angular velocities much smaller than LEOs, making such calibration not accurate enough. Satellites on low orbits that could be used as fast moving calibrators are sparse and less convenient to use due to short observing time windows. Therefore there is a need for a method that would provide optical calibration signals with an accuracy at the level of 0.1 millisecond with respect to UTC timescale in a stable and convenient manner for image timing error measurements.

Low-orbit artificial satellites astrometry, as well as observing stellar occultations by small 
 and fast moving asteroids, poses an observational challenge connected with the image timing accuracy. 
 Occultation events often have duration of less than one second (e.g. 0.218 seconds in case of occultation by Phaethon from 15 October 2019, ACROSS campaign),
 and for scientifically usable and consistent results between observers, the timing precision needs to be of the order of milliseconds. 
 The GNSS-based (Global Navigation Satellite System) time stamps enable such precision, 
 the problem are unknown camera delays, shutter delays and other instrumental effects, deteriorating the timing. 
 Thus the fast observing systems and cameras used for such observations should be tested for their timing precision 
 and possible delays, with an order of magnitude better time resolution. 
 So far, probably the only device enabling to test camera timing accuracy has been SEXTA (Southern EXposure Timing Array), described 
 by \cite{Barry2015}, constructed basing on the original idea of EXTA (EXposure Time Analyser) by \cite{dangl}. The devices consist of an array of blinking diodes, 
 precisely timed by a GNSS receiver, to be imaged by scientific camera under study, in order to compare
 the template timing from GNSS (coded as a set of diodes that are on) with the camera internal time stamp, usually saved in the FITS image header. 
 SEXTA array allows temporal resolution down to 2 ms (milliseconds).
 There is also a software tool implemented in Cyanogen Imaging MaximDL, called Shutter Latency Measurement. It displays a special sequence of images in a PC monitor which when recorded by a camera allows to measure timing errors. Unfortunately this solution depends on PC internal delays and is precise to only 10 ms.
 New observational challenges, however, require much 
 better timing resolution, thus the idea, design (by KK) and successful construction of NEXTA (New EXposure Timing Analyser).
 
 In this work we describe NEXTA simple design, possible to be reproduced by 
 anyone interested using inexpensive and simple components. The instructions, including wiring scheme are added in the Appendix \ref{App. A}. The operating code is available from our Institute server, see Appendix \ref{App. B}. This work describes also the possibilities and limitations of NEXTA deduced from extended testing  on a range of astronomical cameras commonly used for stellar occultations (Sect. \ref{Sect. 2}), assessing their usability for such observations. Section \ref{Sect. 3} presents the testing results for cameras used for observations of artificial satellites. The last section summarises the NEXTA suitability and camera testing results.

\section{Construction and tests of NEXTA on occultation cameras}
\label{Sect. 2}
In this section we discuss NEXTA from the point of view of observing stellar occultations by small bodies. Favored by strongly improved predictions (Gaia EDR3, \cite{Gaia_EDR3}) and the provision of modern CMOS cameras, the current trend in occultation astronomy is towards ever smaller and/or closer-to-Earth objects with correspondingly increasing requirements for timing. 

NEXTA is an innovative development and offers a wide range of applications, particularly with regard to testing the timing capabilities of occultation recording devices. In the following we describe how to build and set up the instrument and give examples of its use in occultation astronomy.
\subsection{NEXTA replication}
\label{Sect. 2.1}
%

\begin{figure}
   \centering
   \includegraphics[width=10cm]{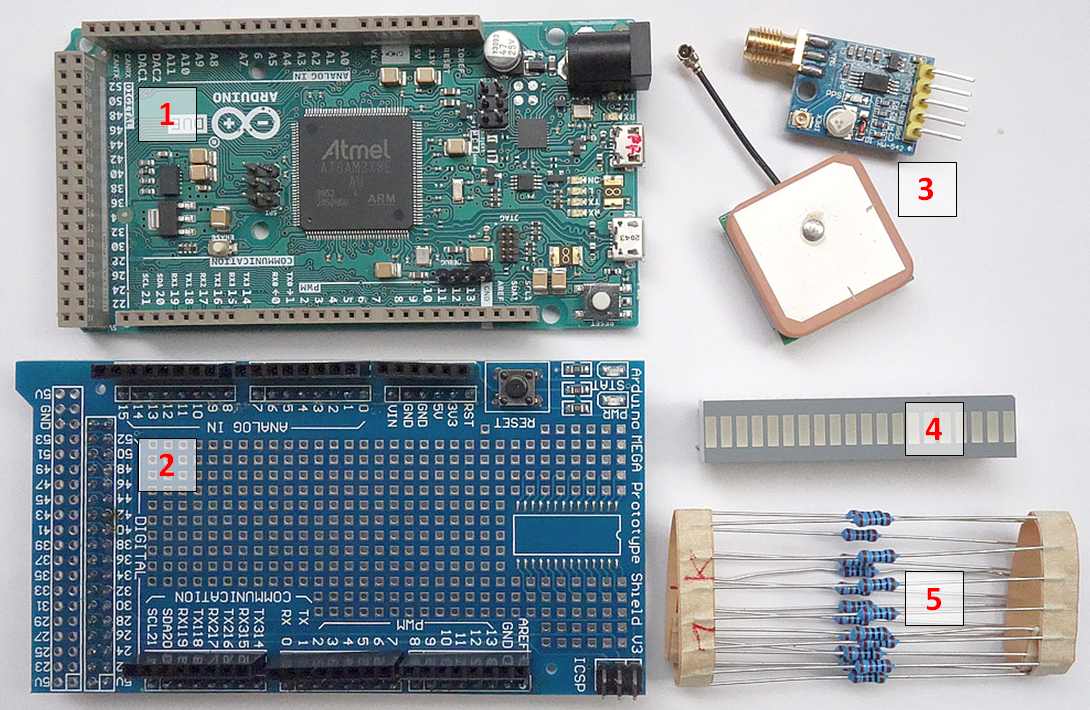}
   \caption{NEXTA main parts. (1) Arduino Due, (2) Arduino Proto Shield, (3) GNSS module with internal antenna and external antenna socket, (4) 20-LED strip, (5) Resistors 1 k Ohm.}
    \label{Fig. 1}
\end{figure}

\begin{figure}
   \centering
   \includegraphics[width=10cm]{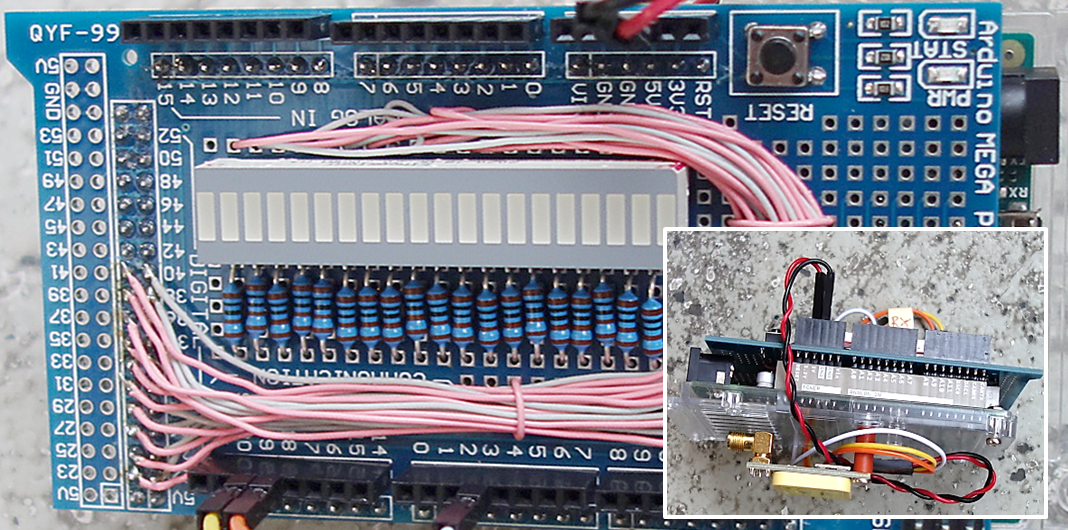}
   \caption{NEXTA completed. Front view with 20-LED display. \textit{Bottom right:} View from above, GNSS module at the bottom of the image.}
    \label{Fig. 2}
\end{figure}

\begin{figure}
   \centering
   \includegraphics[width=10cm]{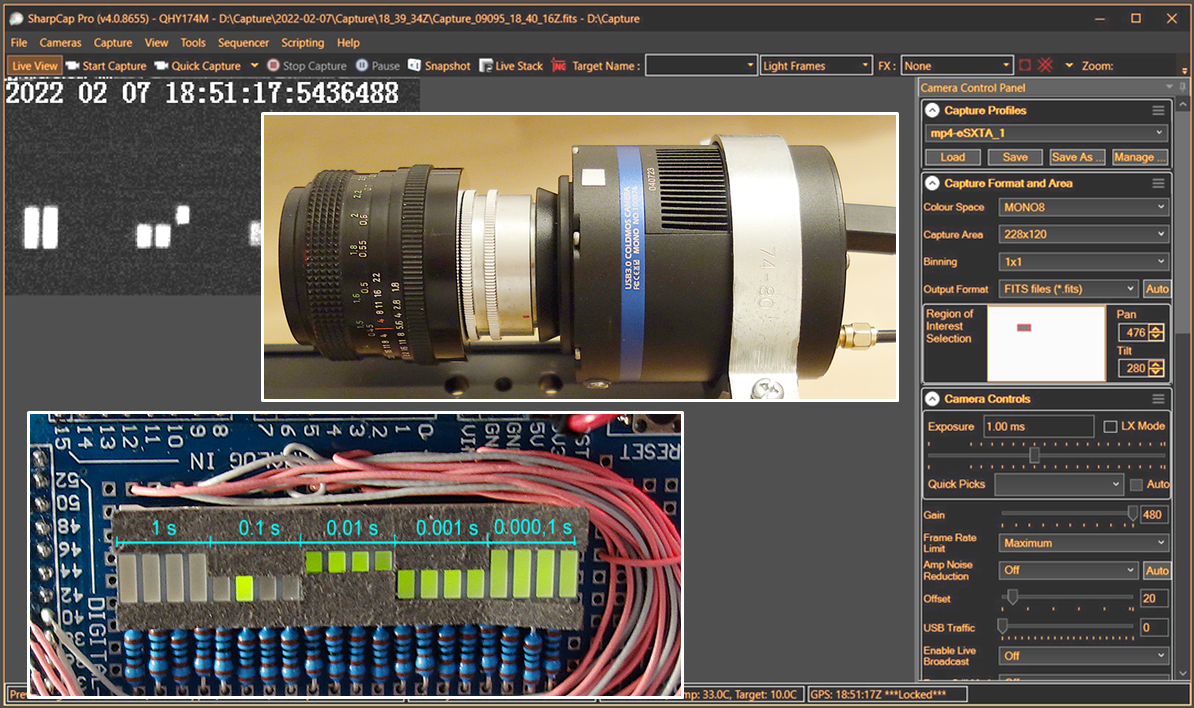}
   \caption{Test setup example. SharpCap 32bit Pro version 4.0.8655.0 recording software, NEXTA display (equipped with a mask for better differentiation of the LED sections, indicated in cyan. The current LED status intentionally does not match the SharpCap display.), QHY174M-GPS camera with a photo lens.}
   \label{Fig. 3}
\end{figure}

\begin{figure}
   \centering
   \includegraphics[width=10cm]{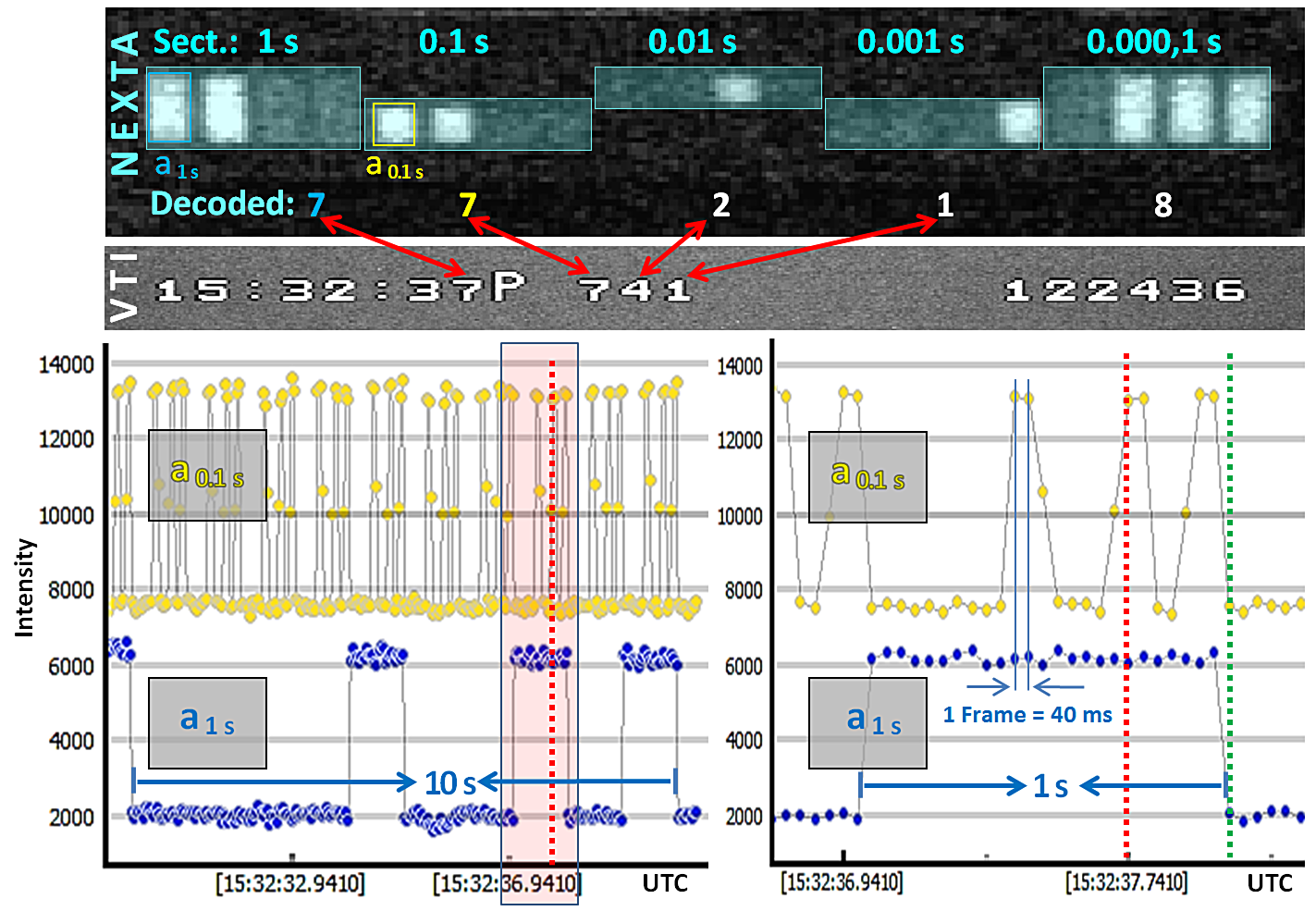}
   \caption{Exemplary demonstration of the functionality of NEXTA using the WAT-910HX-RC camera / VTI recording system described in Sect. \ref{Sect. 2.4.1}. Top image shows a 20 ms half-frame (= 1 video field), field number 122436, of an AVI recording of NEXTA's display, cyan: LED sections (digits), 'Decoded': Digits decoding according to Fig. \ref{Fig. 5}, a\textsubscript{1s} and a\textsubscript{0.1s} are the LEDs photometrically read by PyMovie version 3.3.2. The corresponding diagrams ("light curves", LED intensity over VTI generated UTC timestamps) from PyOTE version 4.6.4 are shown below, where the one on the right represents a temporal stretching of the area marked in red on the left. The yellow 0.1 s digit curves are shifted up in intensity by 6000 units for better visibility. The red dotted vertical lines correspond to the analogue time (15:32:37.741 UTC) indicated on the video field above. The green dotted line indicates the end of the 15:32 37th UTC second, both LEDs went out at the same time. This and the identically shaped curves (with for the a-LED positions exactly 3 phases with LED ON) show the synchronous working mode of the NEXTA sections 1 s and 0.1 s. The three remaining sections work in the same way. The exposure time was 10 $\mu$s, so even the sections below 0.1 s show individual LEDs lit. The "VTI" marked area shows the VTI generated time mark stamped into the video field. The red arrows show the correspondences of the digits. The 20 ms time difference within the 0.01 s digit is due to an instrumental delay of the camera, for details see Sect. \ref{Sect. 2.4.1}}
    \label{Fig. 4}.
\end{figure}

\begin{figure}
   \centering
   \includegraphics[width=5.5cm]{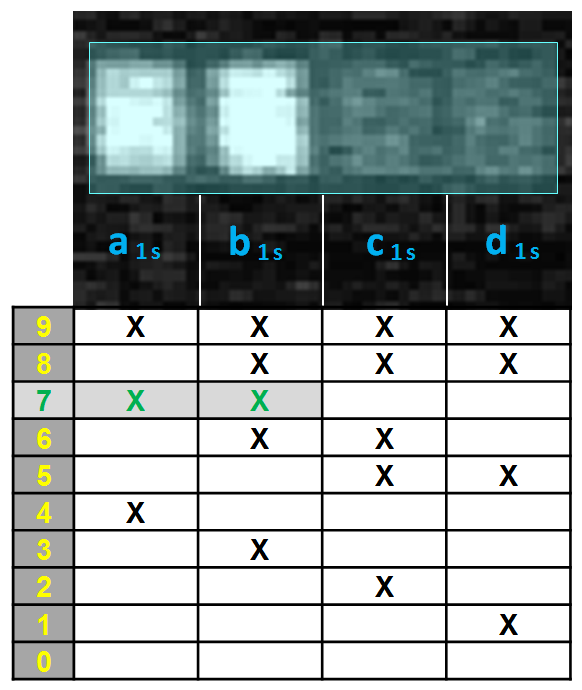}
   \caption{Decoding of NEXTA's LED locations (marked in blue) into analogue numbers (yellow), shown for the 1 s LED section. X = LED ON, no sign = LED OFF, green marks refer to the LED's state presented on the top. This scheme is valid for all of the five sections.}
    \label{Fig. 5}
\end{figure}

\begin{figure}
   \centering
   \includegraphics[width=12cm]{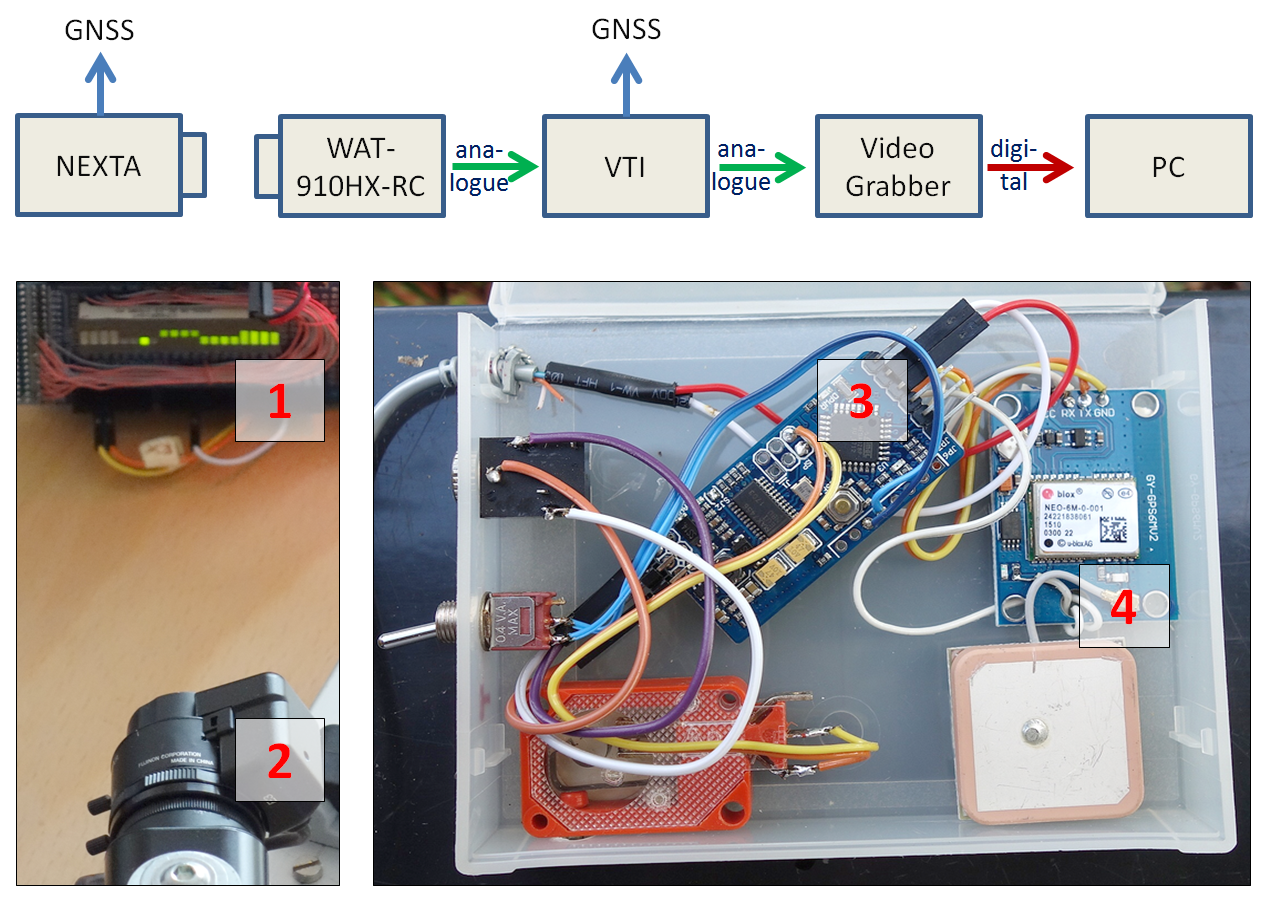}
   \caption{NEXTA / WAT-910HX-RC / VTI test setup. (1) NEXTA, (2) WAT-910HX-RC with lens; VTI interior view: (3) Arduino Uno, (4) GNSS module with antenna.}
    \label{Fig. 6}
\end{figure}

\begin{figure}
   \centering
   \includegraphics[width=10cm]{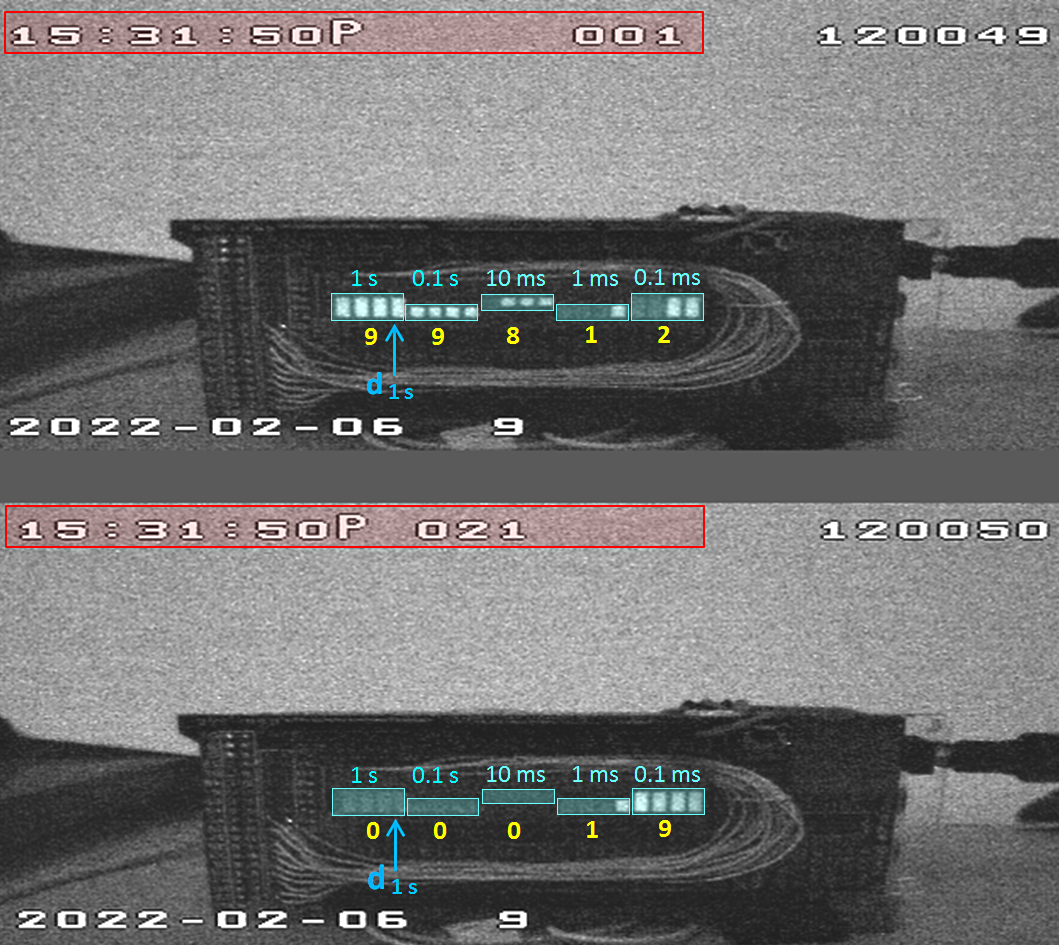}
   \caption{Two 20 ms duration consecutive video fields of a WAT-910HX-RC 25 FPS PAL video (see  Sect. \ref{Sect. 2.4.1}). Red marked: Time stamps imprinted on the video stream by the VTI [h:min:s  ms]. Cyan: NEXTA sections. Yellow: NEXTA's decoded analogue UTC time stamps [s]. Blue: For light curve simulation (Fig. \ref{Fig. 8}) used LED location d\textsubscript{1s} of NEXTA's 1 s digit section.}
   \label{Fig. 7}
\end{figure}   

\begin{figure}
   \centering
   \includegraphics[width=10cm]{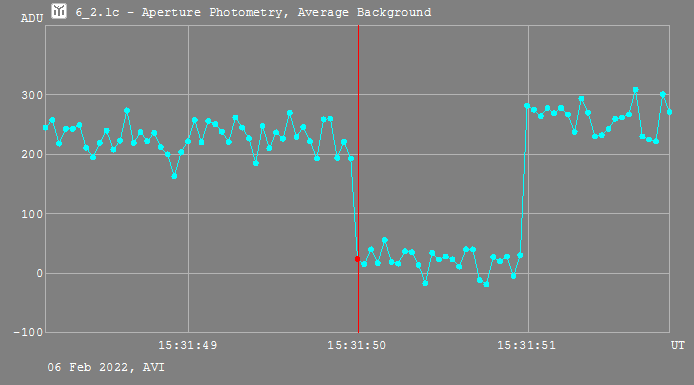}
   \caption{Tangra light curve of the state of the d\textsubscript{1s} LED (see Fig. \ref{Fig. 7} and Table \ref{Tbl. 1}) obtained from the test video described in Sect.  \ref{Sect. 2.4.1}. The red line refers to frame 127, video field 120050, as shown in Table \ref{Tbl. 1}.}
   \label{Fig. 8}
\end{figure}

\begin{figure}
   \centering
   \includegraphics[width=10cm]{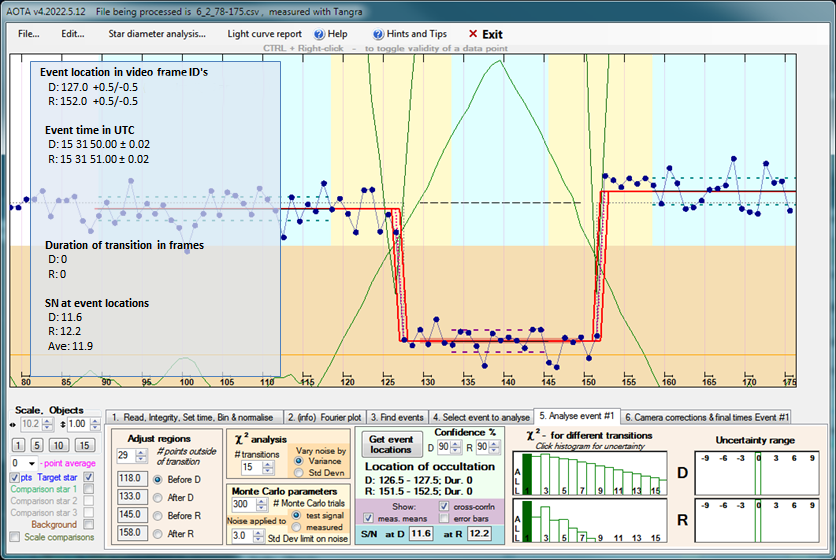}
   \caption{AOTA analysis of the light curve shown in Fig. \ref{Fig. 8}}
   \label{Fig. 9}
\end{figure}

\begin{figure}
   \centering
   \includegraphics[width=10cm]{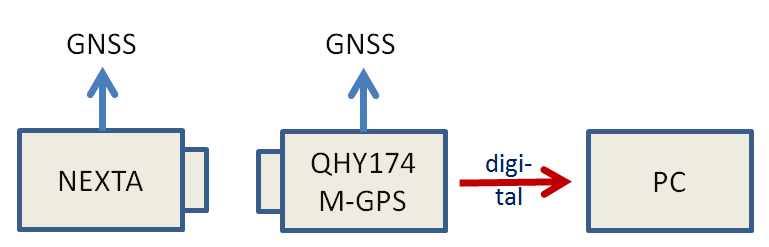}
   \caption{Test setup for the NEXTA tests with the QHY174M-GPS camera.}
   \label{Fig. 10}
\end{figure}

\begin{figure}
   \centering
   \includegraphics[width=10cm]{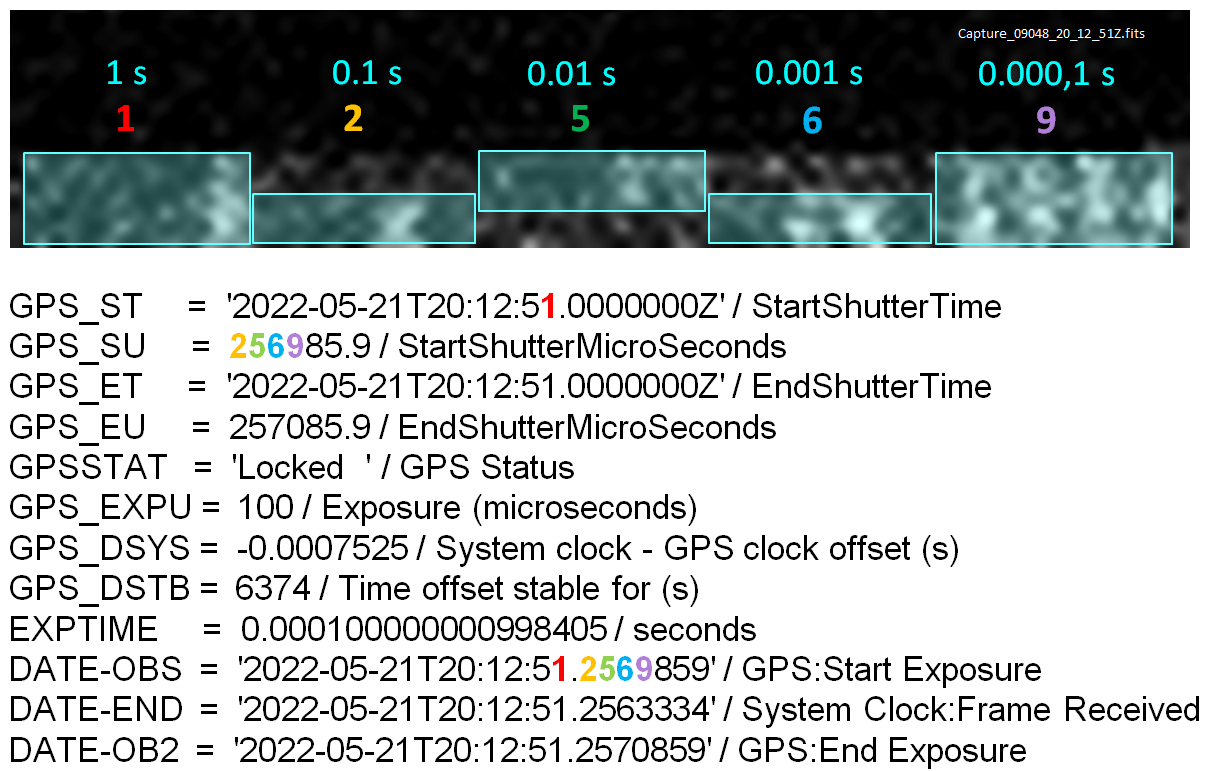}
   \caption{Single QHY174M-GPS camera FITS frame (Capture-09048-20-12-51Z.fits) time-relevant FITS keywords recorded with SharpCap (recording details see text). On the top there is imaged the NEXTA display and (coloured) decoding of its digits, indicating the start of a 100\mbox{ $\mu$s} exposure, identically represented by the FITS keywords GPS\_ST, GPS\_SU and DATE-OBS. The FITS keywords GPS\_ET, GPS\_EU and DATE-OB2 relate to the end of exposure.}
   \label{Fig. 11}
\end{figure}

\begin{figure}
   \centering
   \includegraphics[width=10cm]{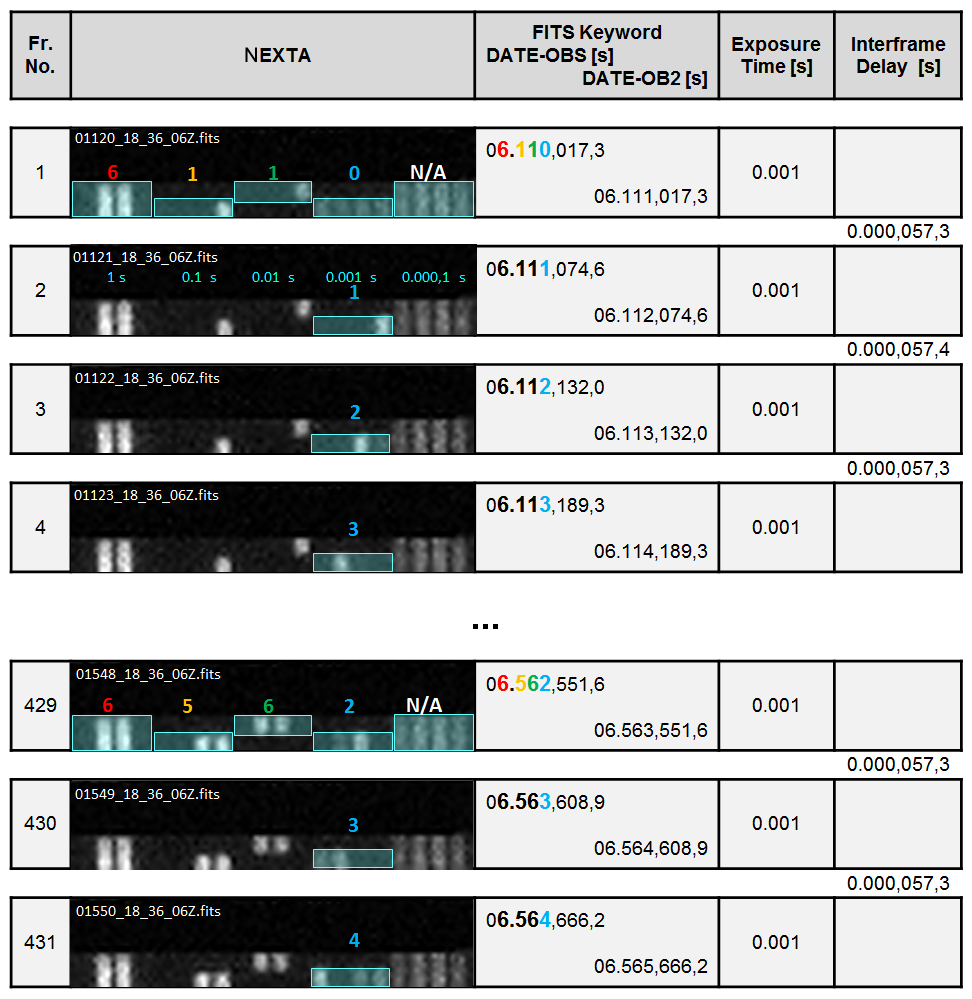}
   \caption{First 4 and last 3 frames of a 431 dropped frame free section from a 30 s FITS sequence taken with a QHY174M-GPS camera capturing the NEXTA display with 1 ms exposure time. DATE-OBS is the SharpCap written FITS keyword representing the GNSS-PPS controlled start of frame in seconds after 2022-05-21T18:36 UTC. DATE-OB2 is the corresponding keyword for the end of the 1 ms exposure, which is followed by an interframe delay of 57.3 $\mu$s. The colored numbers show the correspondence of the respective decoded NEXTA digits with their counterparts in the FITS keyword DATE-OBS. Using 1 ms exposure time, NEXTA's 0.000,1 s digit cannot be time resolved.}
   \label{Fig. 12}.
\end{figure}

\begin{figure}
   \centering
   \includegraphics[width=10cm]{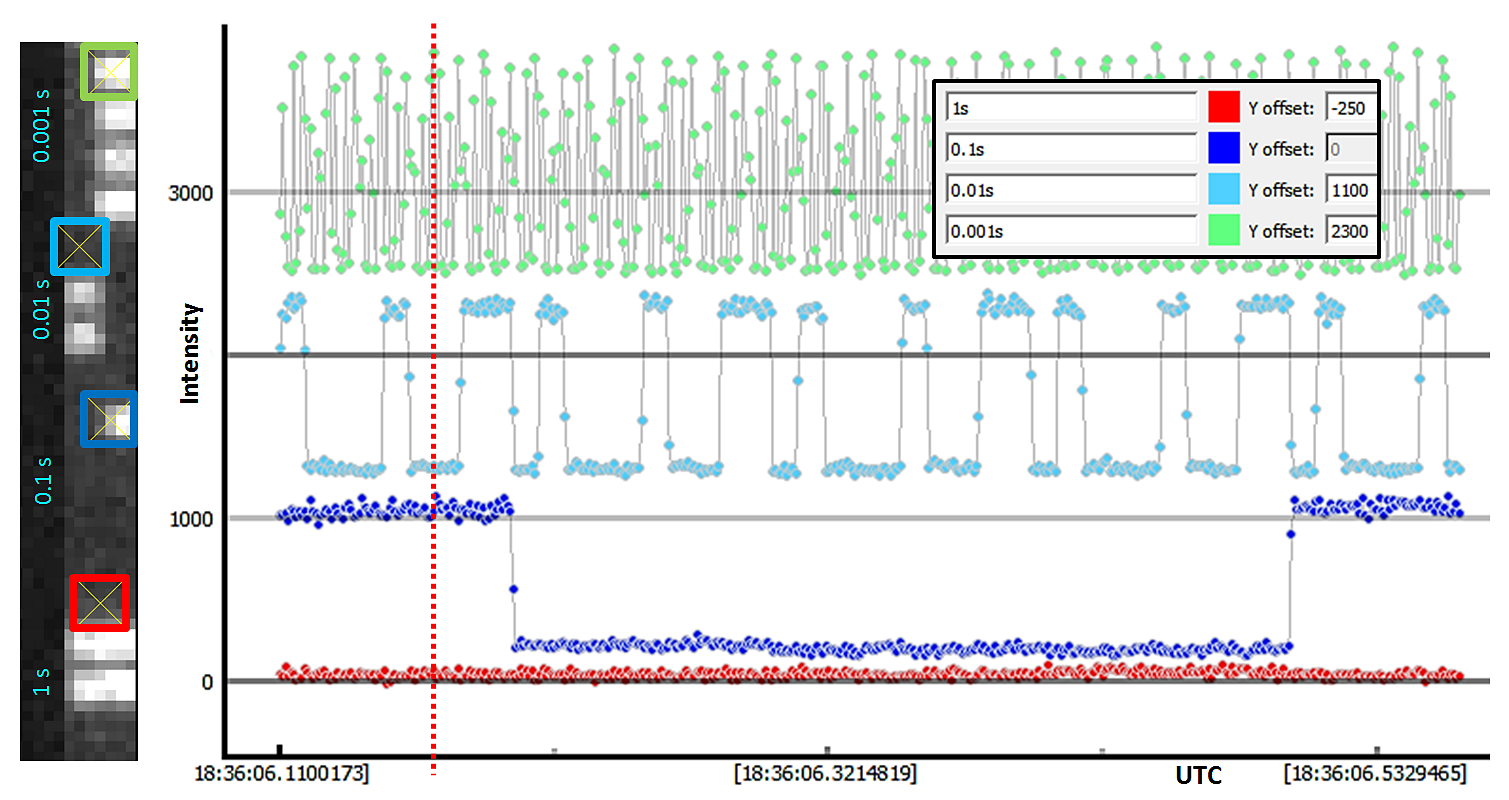}
   \caption{QHY174M-GPS dropped frame free FITS sequence of 431 1-ms frames imaging NEXTA's 1 s to 0.001 s digits (on the left). The digit 0.000,1 s is omitted because there is no time resolution. The curves represent the PyMovie read out d-LED states versus the GNSS controlled time form the camera, plotted with PyOTE. The curve colors correspond to those of the d-LED positions on the left. For better visibility, the curves are shifted vertically by an amount seen from the inline image top right. The state of the 1 s d-LED (= OFF) did not change during the recording (red curve). The red dotted line refers to the current d-LED states on the left side. The blue curve will be considered as occultation light curve, see Fig. \ref{Fig. 14}}
   \label{Fig. 13}
\end{figure}

\begin{figure}
   \centering
   \includegraphics[width=10cm]{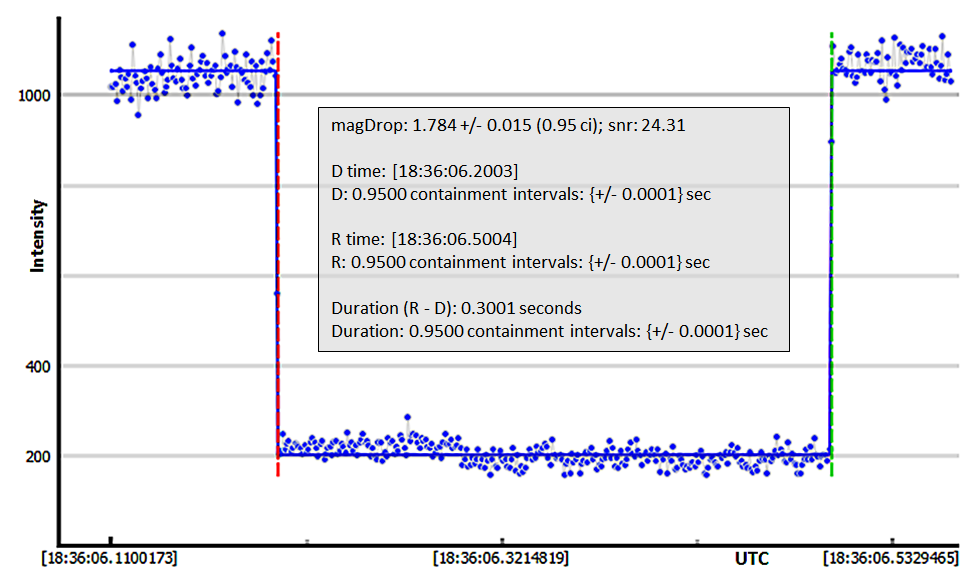}
   \caption{PyOTE solution of a simulated light curve drop, derived from the 0.1 s digit of NEXTA (see the curve in blue in Fig. \ref{Fig. 13}).}
   \label{Fig. 14}
\end{figure}

\begin{figure}
   \centering
   \includegraphics[width=10cm]{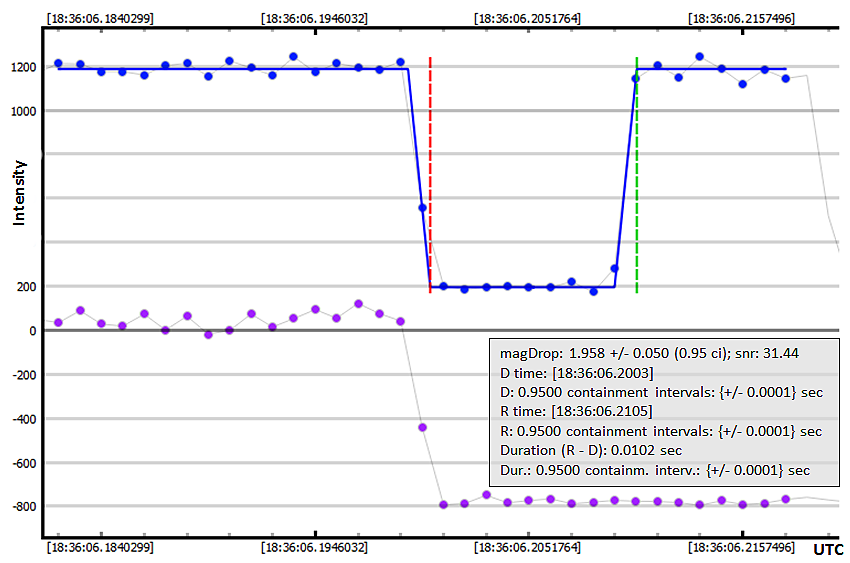}
   \caption{PyOTE solution of the 0.01 s digit light curve (here blue, light blue in Fig. \ref{Fig. 13}); violet: 0.1 s digit light curve (in Figs. \ref{Fig. 13} and \ref{Fig. 14} blue). The D time here agrees with the D time in Fig. \ref{Fig. 14}.}
    \label{Fig. 15}
\end{figure}

\begin{figure}
   \centering
   \includegraphics[width=10cm]{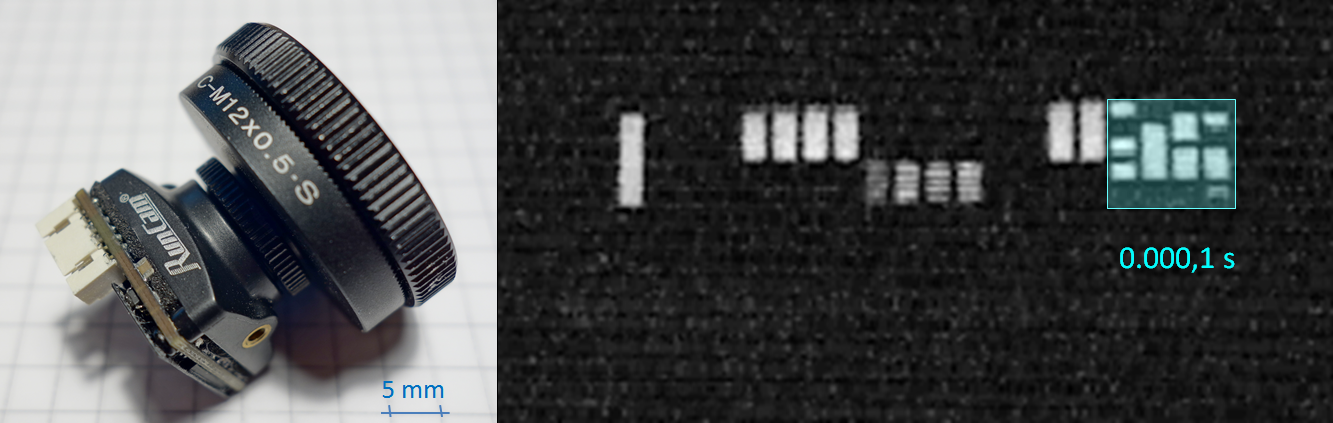}
   \caption{NEXTA rolling shutter detection on a RunCam Night Eagle 3 camera. On the left, the camera equipped with a C-mount adapter; on the right, a single frame (40 ms duration with exposure time 0.039 ms) showing typical rolling shutter effects in the 0.000,1 s digit of NEXTA.}
    \label{Fig. 16}
\end{figure}

\begin{figure}
   \centering
   \includegraphics[width=10cm]{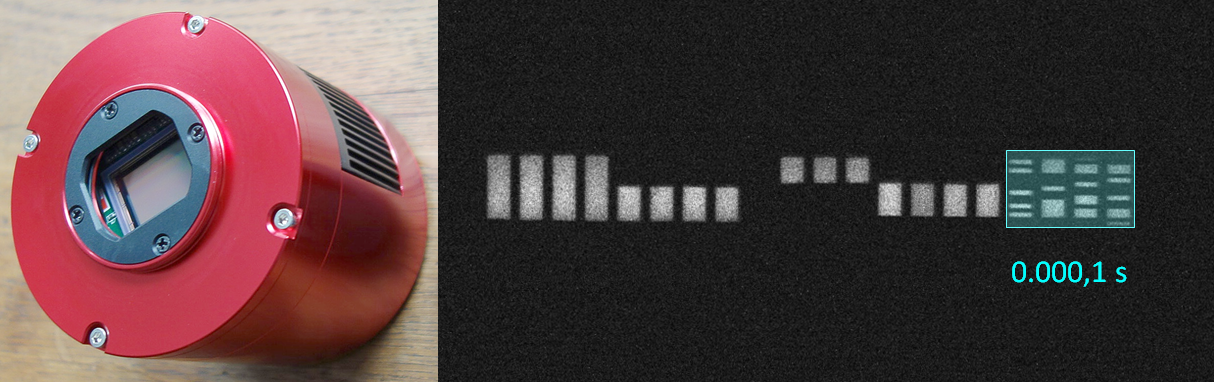}
   \caption{NEXTA rolling shutter detection on a ZWO ASI1600MM camera. On the left, the camera (1.8/50mm photo lens, used for the test, removed); on the right, a single frame (exposure time 50 $\mu$s) showing typical rolling shutter effects in the 0.000,1 s digit of NEXTA.}
    \label{Fig. 17}
\end{figure}

\begin{figure}
   \centering
   \includegraphics[width=10cm]{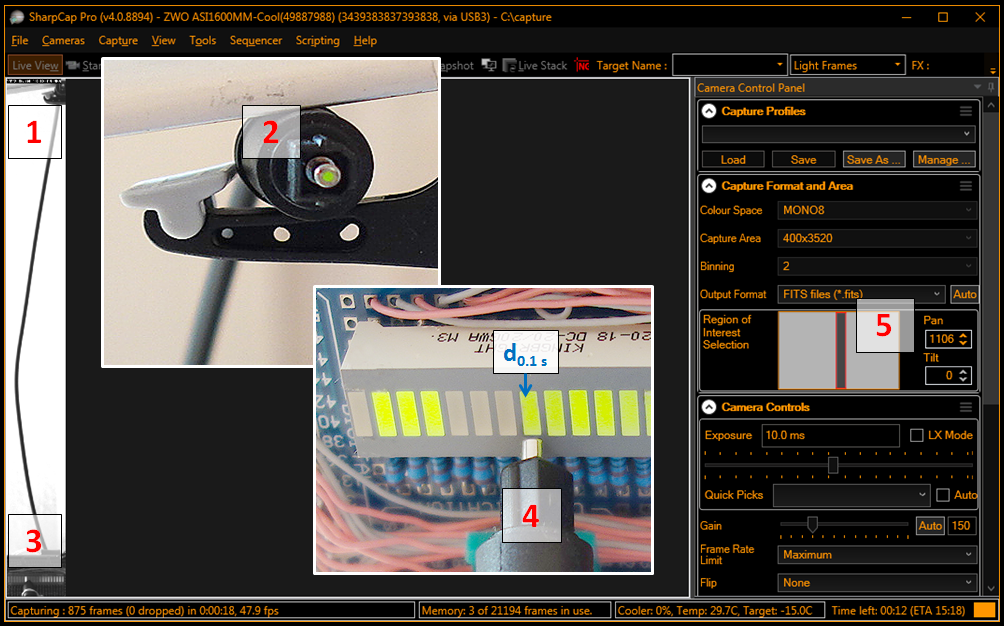}
   \caption{Test setup to determine the rolling shutter delay of a ZWO ASI1600MM camera. In the background the SharpCap user interface with the sensor region of interest (5) used and the frame live view (1),~(3). (4),~(3): Fibre optic cable entrance imaging NEXTA's d\textsubscript{0.1s} LED. (2),~(1): Fibre optic cable output.}
   \label{Fig. 18}
\end{figure}

\begin{figure}
   \centering
   \includegraphics[width=10cm]{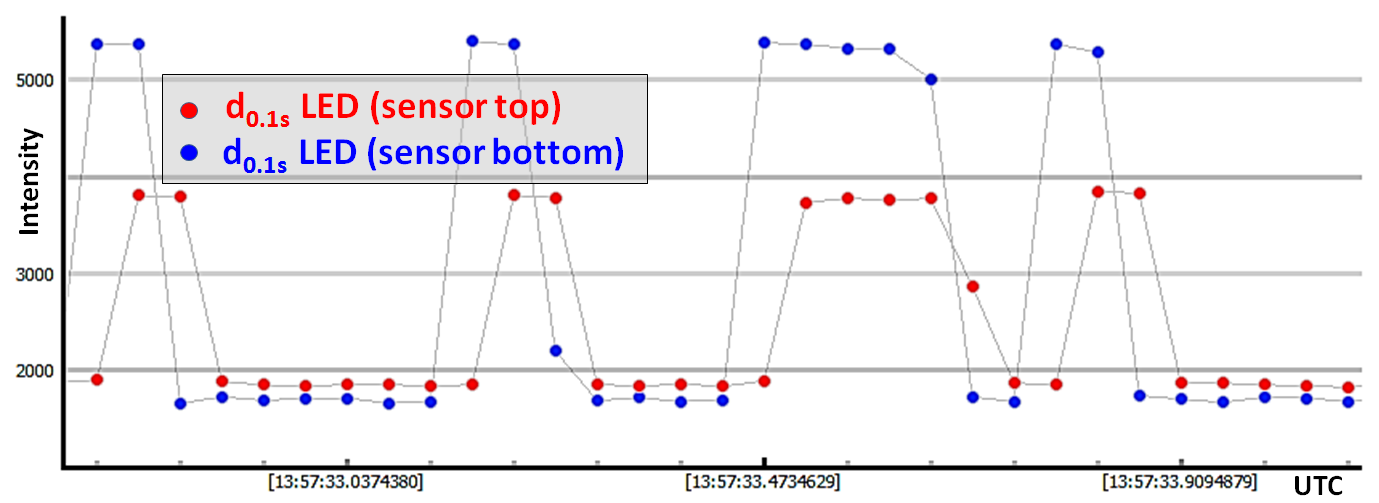}
   \caption{ZWO ASI1600MM camera rolling shutter delay test. For the experimental setup see Sect. \ref{Sect. 2.3.3} and Fig. \ref{Fig. 18}. Both curves represent the state of NEXTA's d\textsubscript{0.1s} LED. The red dotted curve is from the fibre optic cable output mapped to the top of the sensor. Due the effect of the rolling shutter, this curve is shifted in time by 1 frame (44 ms) compared to the blue dotted curve from NEXTA's d\textsubscript{0.1s} LED directly mapped to the bottom of the sensor. The test case showed here is without binning.}
   \label{Fig. 19}
\end{figure}

\begin{figure}
   \centering
   \includegraphics[width=10cm]{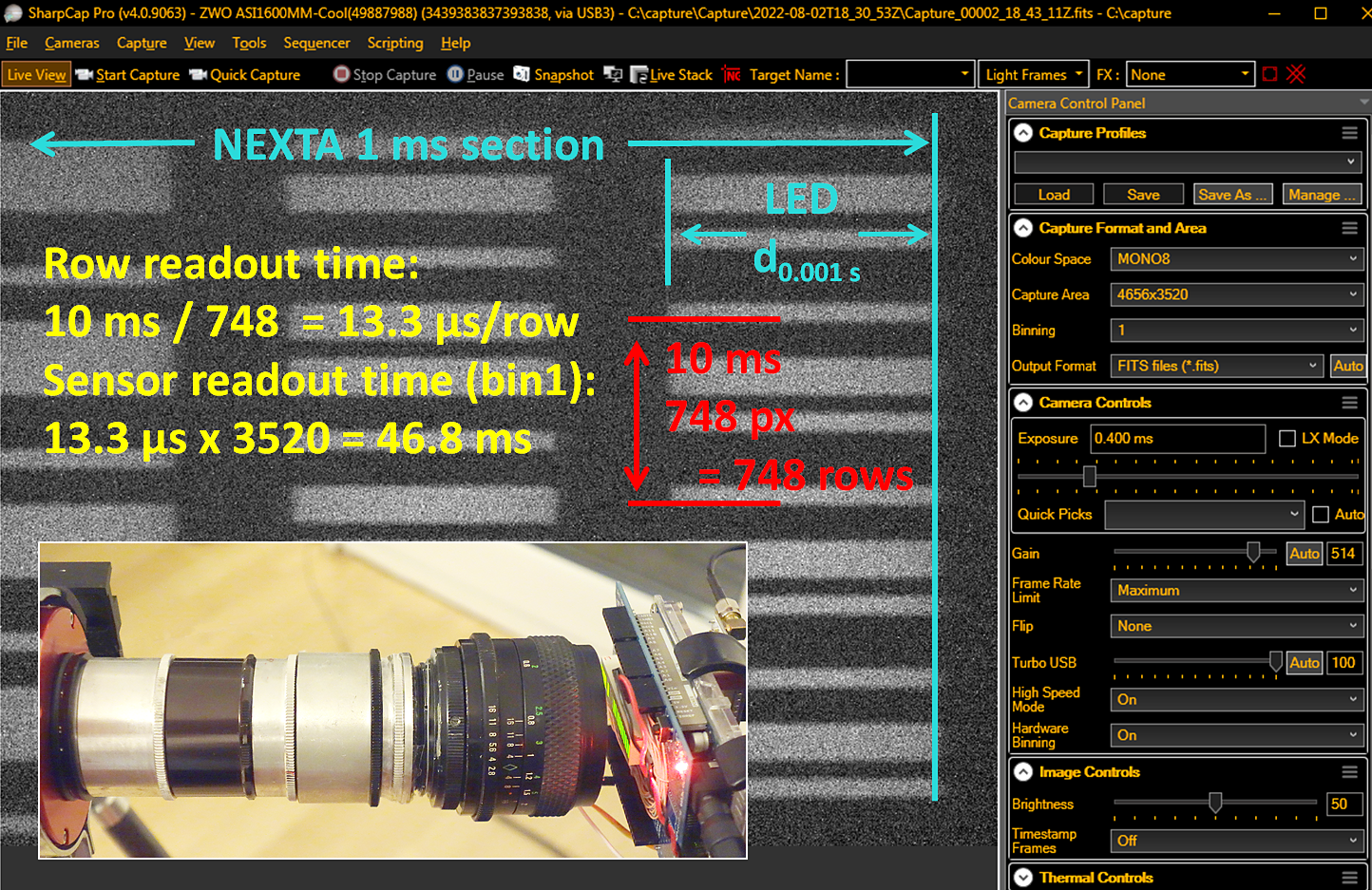}
   \caption{NEXTA ZWO ASI1600MM rolling shutter delay measurement. In the background SharpCap, capturing parts of NEXTA's 0.001 s digit (exposure time 0.4 ms). The inserted image shows the setup with spacer rings and a 2.8/35 mm photo lens for imaging the  d\textsubscript{0.001s} LED onto the camera sensor. From the light pattern (compare Fig. \ref {Fig. 5}) of the d\textsubscript{0.001s} LED a row readout time of 13.3 µs was measured, corresponding to 46.8 ms for the entire sensor.}
   \label{Fig. 20}
\end{figure}

\begin{figure}
   \centering
   \includegraphics[width=10cm]{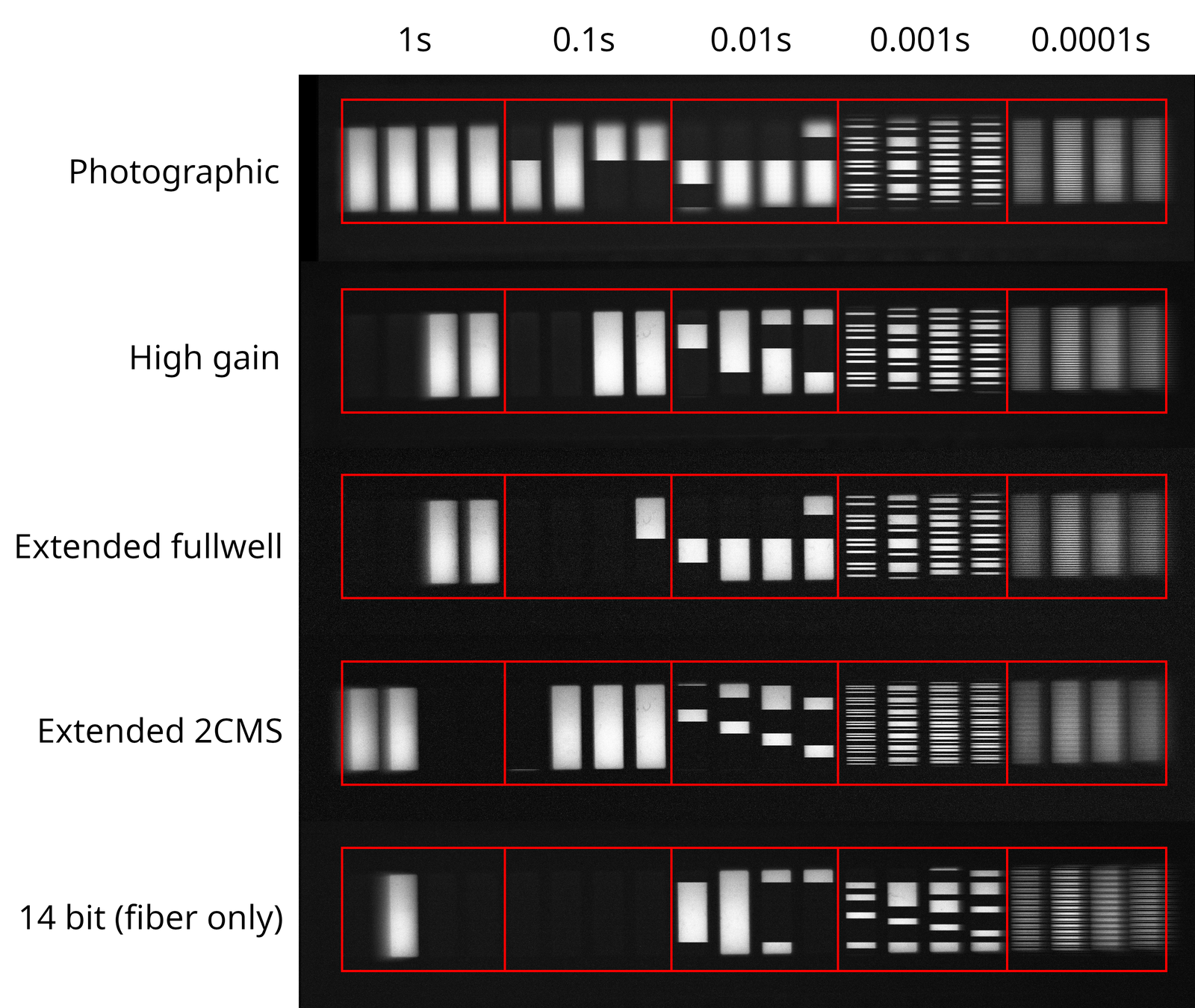}
   \caption{Example images of the NEXTA recorded with QHY 600M Pro camera during tests in 5 different modes described in Sect. \ref{Sect. 3.1}. Rolling shutter effect is best visible in three sections with largest frequency of blinking (0.01s, 0.001s, 0.0001s). Such data was used to calculate the delay between readout of consecutive rows and subsequently to calculate the UTC timing of the first row of the camera sensor.}
   \label{Fig. 21}
\end{figure}

\begin{figure}
   \centering
   \includegraphics[width=10cm]{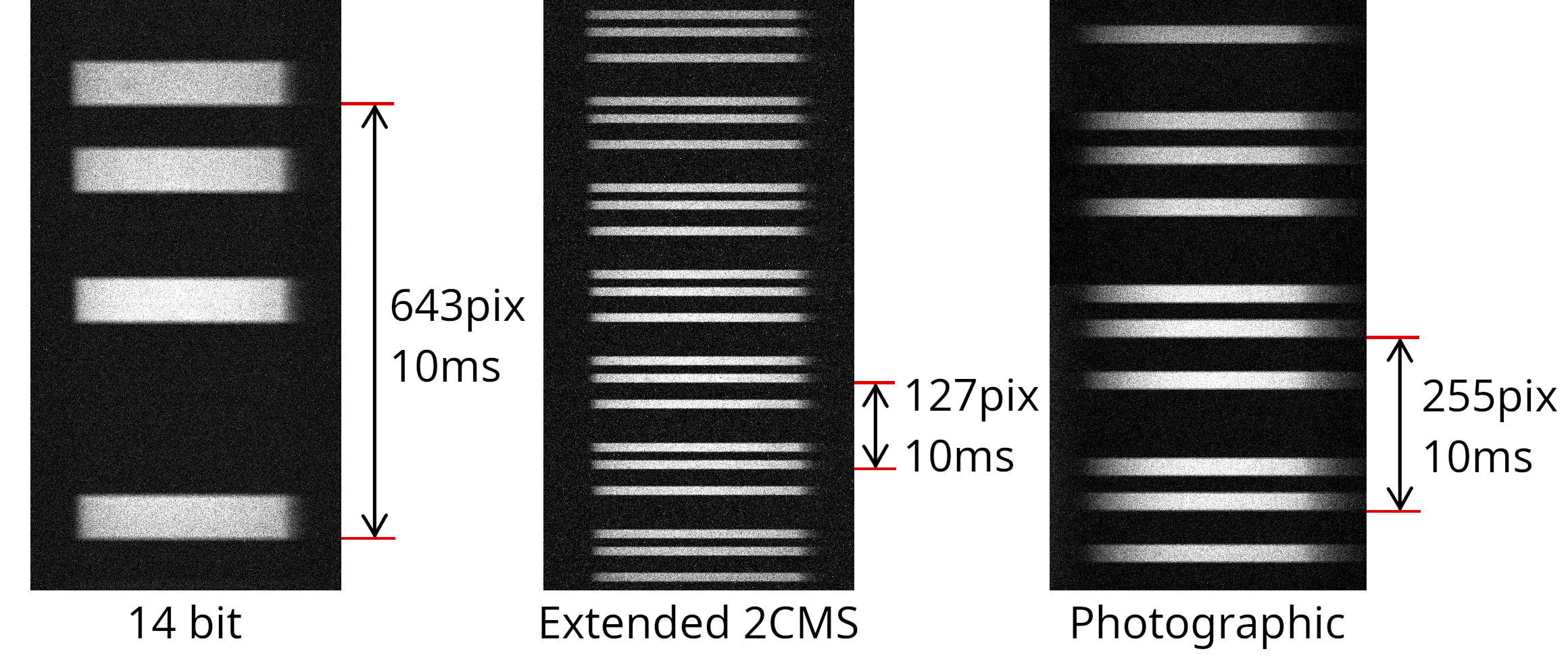}
   \caption{Subsection of single images of NEXTA with exposure time of 70$\mu$s showing the same single LED ($a_{0.001s}$) recorded with QHY 600M Pro camera during tests described in Sect. \ref{Sect. 3}. Due to rolling shutter working with different readout speeds (14bit mode - 15.6$\mu$s/line, Extended 2CMS mode - 78.1$\mu$s/line, Photographic mode - 39.1$\mu$s/line) different recordings of the exact same sequence of LED blinking are visible. By dividing the time between selected blinks of the LED with the number of pixel rows at which they were recorded we derived single row readout time of this camera (see Table \ref{Tbl. 2}).}
   \label{Fig. 22}
\end{figure}

\begin{figure}
   \centering
   \includegraphics[width=10cm]{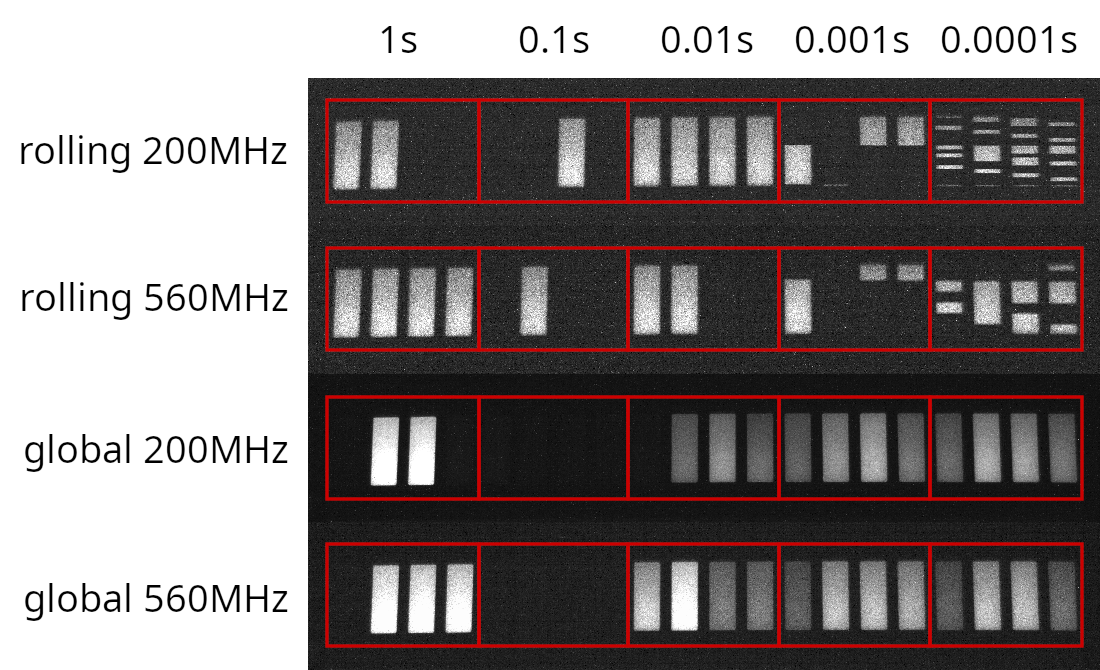}
   \caption{Example images of the NEXTA recorded with Andor Zyla 5.5 camera during tests in 4 different modes described in Sect. \ref{Sect. 3.2}. Exposure time was always the same - 27$\mu$s. Top images were taken using rolling shutter, bottom images - global shutter. In global shutter mode the sections of NEXTA blinking at 1ms and 0.1ms intervals were always recorded with 4 LED on, even though in rolling shutter mode the same sections are recorded with changing LED state without problems. The images taken in global shutter mode are also significantly better exposed. It looks like they were acquired with about 10ms exposure time, even though it was set in software and reported in FITS header as 27$\mu$s.}
   \label{Fig. 23}
\end{figure}

\begin{figure}
   \centering
   \includegraphics[width=10cm]{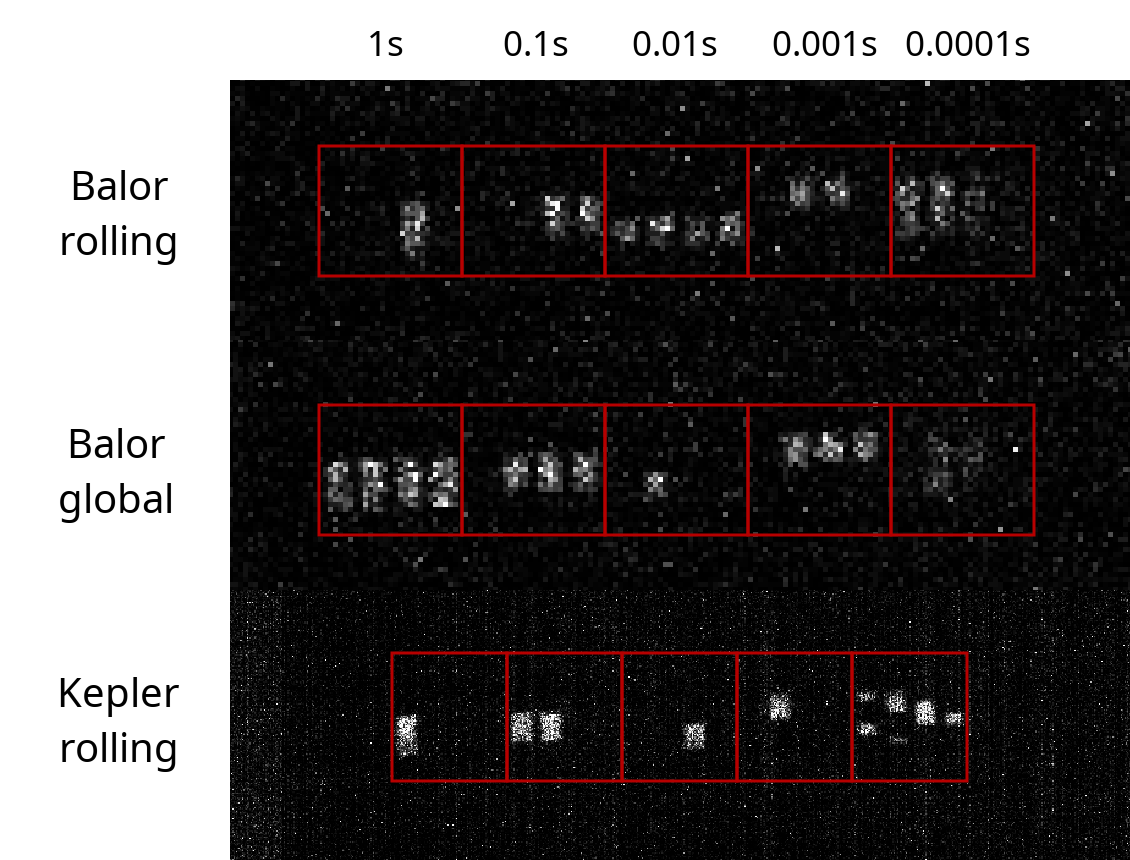}
   \caption{Example images of the NEXTA recorded with Andor Balor 17-12 and FLI Kepler 4040, described in Sect. \ref{Sect. 3.3} and Sect. \ref{Sect. 3.4}. }
   \label{Fig. 24}
\end{figure}


All the necessary parts to build NEXTA are available off the shelf at a reasonable price. Figure \ref{Fig. 1} shows the main components. A wiring diagram (Fig. \ref{Fig. 24}) as well as some special hints are given in Appendix \ref{App. A}. There is extensive freedom in the specific design of the device. This makes it easily adaptable to different test setups. Sufficient computing power must be ensured for the micro-controller. Our tests show that Arduino Mega controller board (16MHz 8-bit Atmel processor) is insufficient to achieve the 0.1ms resolution, while Arduino Due (84MHz 32-bit ARM processor) is perfectly adequate for the task.  The NEXTA prototype presented here has an internal GNSS antenna and also a socket for connecting an external antenna. With the latter, the device can be placed indoors for more defined optical conditions. Figure \ref{Fig. 2} shows the built prototype without housing.

After completion the device has to be programmed using Arduino Integrated Development Environment (IDE)\footnote[1]{\url{https://www.arduino.cc/en/software}}. The NEXTA software utilises hardware interrupts to ensure the lowest possible internal latency. Each GNSS impulse (which is setup to PP10S - one pulse per 10 seconds) triggers an unambiguous sequence of LEDs blinking at a frequency of 10kHz for 10 seconds. If no further impulses arrive the device displays 3 LEDs ($a_{1s}$, $c_{1s}$ and $a_{0.1s}$) constantly on as an indication for the user. This combination is never displayed during any other situation.
The LEDs timing between PP10S impulses is based on the 12MHz oscillator on the Arduino board.
The error of this oscillator is measured with respect to the GNSS signals after each startup. Than if drift calibration is selected in source code (code line 17) it is used during the subsequent operations of NEXTA. By default the drift it only is compared with limit. If it is too large (the default limit is 10 microsec / sec, configurable in code line 20) the device does not run and two LEDs ($a_{1s}$ and $c_{1s}$) are constantly on as an error message. During our tests we were always below that limit, but clock drift may be different for each Arduino, and may change with temperature. During initial setup the device displays 4 or more LEDs constantly on as an indication for the user about the setup progress (see \ref{App. C} for details).
The device can be powered by 5~V (USB socket) or 12 V. If GNSS satellite signals can be received in sufficient quality, the device is in stable operation after about 10~minutes. A typical test setup is shown in Fig. \ref{Fig. 3}, using SharpCap\footnote[1]{\url{https://www.sharpcap.co.uk/}} as capture software and a QHY174M-GPS\footnote[2]{\url{https://www.qhyccd.com/qhy174gps-imx174-scientific-cooled-camera/}} camera.
\subsection{NEXTA mode of operation}
\label{Sect. 2.2}
NEXTA provides 5 analogue UTC digits (sections), ranging from 1~s to 0.1~ms (Fig. \ref{Fig. 3}). Each digit consists of 4 LEDs controlled by the GNSS receiver's PP10S signal, which has an accuracy several orders of magnitude higher than the 0.1 ms temporal resolution of the NEXTA. The NEXTA mode of operation is shown in an example, presented in Fig. \ref{Fig. 4} (software used: PyMovie\footnote[1]{\url{https://pypi.org/project/pymovie/}} and PyOTE\footnote[2]{\url{https://pypi.org/project/pyote/}}). 

The decoding of NEXTA's display sections into analogue numbers occurs according to the scheme shown in Fig. \ref{Fig. 5}. The comparison of the times displayed visually by NEXTA with the own time stamps of the devices to be tested results in their temporal accuracy. However, the camera must allow for sufficiently short exposure times, in order to be able to use the highest resolution 0.000,1 s LED section, for example. This requires electronic shutters, ideally global ones. Sensor row readout delays of rolling shutter cameras can also be detected and quantified with NEXTA, see Sects. \ref{Sect. 2.3.3} and \ref{Sect. 3.1}.
Another limitation on the camera side with regard to the utilization of NEXTA's higher temporal resolutions are timing errors, for example as a result of dropped or backwards jumping frames. Such errors will increase when the frame rate (FPS, frames per second) of the recording system reaches its limit.

The use of common occultation photometry software such as PyMovie and Tangra\footnote[1]{\url{http://www.hristopavlov.net/Tangra/Tangra.html}} facilitates the evaluation although currently NEXTA's visual time stamps have to be decoded manually. The development of a special automatic NEXTA decoding software would, for example, greatly simplify long-term tests of the temporal and thermal stability of test objects.
\subsection{Temporal accuracy of NEXTA}
\label{Sect. 2.3}
The primary functional requirement of NEXTA is that it will display a sequence of LED blinks which is synchronised with UTC time scale. In order to verify that there are no delays larger than 0.1 ms in the displayed LED sequence an experiment was performed using a very high FPS camera ($\sim 10$ kHz) - Andor Zyla 5.5. No dropped frames were reported by Andor Solis camera controll software. NEXTA was recorded together with a single LED which was directly connected to a PPS signal of an additional GNSS receiver. Inspecting individual frames of the recording we found that both NEXTA diodes and GNSS diode displayed a whole second mark at exactly the same time. Since the PPS signal is synchronised with UTC with accuracy of tens of nanoseconds we concluded that there is no measurable delay in time displayed by NEXTA during its operation.

For testing purposes two commands were added to Arduino program. First at the beginning of main loop to change selected digital pin state to high, second at the end of main loop to change that pin state to low. With oscilloscope it was measured that the loop execution time is below 57 microsec while changing LEDs state and below 3 microsec while the LEDs state requires no change. This results show that the overall latency of LEDs display should be adequate for the task, especially because the control program is setting up the highest frequency LEDs first (within 18 microsec from the loop beginning), and the lowest frequency LEDs last.

A 30 hour long-term test was also performed using the global shutter QHY174-GPS camera as a reference. Both NEXTA and the camera were equipped with their own external antennas to ensure good GPS reception (according to the camera's GPS log, there were always 9-12 satellites in sight). SharpCap was used to record 1ms exposure 5 frames FITS sequences every 20 minutes. While the resolution of this test was limited to 1 millisecond it showed no problems with NEXTA readings and perfect agreement with camera timing recorded in DATE-OBS keyword of FITS images.
\subsection{NEXTA typical applications}
\label{Sect. 2.4}
The following is a non-exhaustive description of NEXTA applications in testing stellar occultation equipment and, to some extent, software. Secondly, the tests are also intended to verify the functioning of NEXTA itself. The setup assumes a stable NEXTA GNSS state, the system under test has only to record NEXTA's LED display in an appropriate manner (for an example, see Fig. \ref{Fig. 3}), followed by an analysis of the recording.

\subsubsection{NEXTA for testing a WAT-910HX-RC Camera / VTI occultation recording system}
\label{Sect. 2.4.1}
Before the era of modern CMOS cameras, analogue video cameras were the means of choice for recording stellar occultations. Mainly because of their high sensitivity, these cameras are still used \cite{souami_multi-chord_2020, rommel_stellar_2020}. To time occultation events, video cameras require additional equipment, usually so-called Video Time Inserters (VTIs).
The following NEXTA use case (Fig. \ref{Fig. 6}) involves a WATEC WAT-910HX-RC
\footnote[1]{\url{https://www.watec.co.jp/English/e_mono.html}}
(PAL) video camera, equipped with a lens FUCINON 1:0.95/2.8-8mm, and a VTI (built by CW). The WATEC camera is an interline transfer CCD capable of making simultaneous exposures of all its pixels with exposure time down to 10 $\mu$s. The VTI bases on work of Smolarz\footnote[2]{\url{https://github.com/smopihub/smopiVTI}} and Andre\footnote[3]{\url{https://esop36.de/lections/arduino_vti.pdf}}.
The VTI (Fig. \ref{Fig. 6}) is equipped with a GNSS receiver that provides the PPS signal and an Arduino-controlled unit that imprints UTC-accurate time stamps on the camera's video signal. The VTI does not generate its own delay of the video signal, at least not in the range of the temporal resolution of NEXTA. According to \cite{dangl}, however, the camera shows instrumental delays (time bias), which depend on the settings of the camera.

\vspace{1cm}
\textit{Recording and analysis of a 25 FPS interlaced AVI.}
\vspace{0.25cm}

We used WAT-910HX-RC recordings of the NEXTA display with the camera output overlaid with the VTI time stamps (for recording hardware chain see Fig. \ref{Fig. 6}). The VTI output was digitized using an USB video grabber (Hauppauge USB-Live2\footnote[1]{\url{https://www.hauppauge.de/site/products/data_usblive2.html}}) and recorded as a lossless 25 FPS interlaced AVI using Syntek Video View on a W7-64bit i7 16GB RAM PC.
As a rule, there were no dropped frames during the recordings. Figure \ref{Fig. 7} shows two consecutive fields (half-frames) of a 56 s video taken with an exposure time of 10 $\mu$s due to camera's electronic shutter. In this mode, the camera works without frame integration. The duration of a video field is nevertheless 20 ms (PAL video standard). 

Table \ref{Tbl. 1} presents the analysis of the video field times from Fig. \ref{Fig. 7} and in addition the result from the end (frame 1376) of the 56 s video. Within the 1 ms temporal resolution of the hardware chain under test the time-bias with respect to NEXTA is always below 1 ms.

\begin{table}[]
\caption{Timing analysis of four video fields of the 25 FPS test video.}         
\label{Tbl. 1} 
\centering   
\begin{tabular}{p{4cm} c c c c} 
\hline\hline
\multicolumn{1}{l}{\multirow{2}{*}{Parameter}}        & \multicolumn{4}{c}{Frame No. / Video   field No.}         \\
\multicolumn{1}{l}{}                                    & 126/120049$^a$   & 127/120050$^a$   & 1376/122548  & 1376/122549  \\
\hline
VTI time stamp {[}h:min:s{]}                              & 15:31:50.001 & 15:31:50.021 & 15:32:39.981 & 15:32:40.001 \\
 Camera instrumental delay$^b$ {[}s{]}    & - 0.020      & - 0.020      & - 0.020      & - 0.020      \\
Corrected camera time {[}h:min:s{]}          & 15:31:49.981 & 15:31:50.001 & 15:32:39.961 & 15:32:39.981 \\
NEXTA visual time {[}s{]}                         & 9.981,2          & 0.001,9          & 9.961,7          & 9.981,7         \\
Corrected camera time deviation from NEXTA visual time {[}ms{]} & - 0.2            & - 0.9            & - 0.7            & - 0.7            \\
State of the d\textsubscript{1s} LED of the 1 s digit section     & ON           & OFF          & ON           & ON      \\
\hline    
\end{tabular}
$^a$See Fig. \ref{Fig. 7}, $^b$see Sect. \ref{Sect. 2.4.1}
\end{table}

\vspace{0.5cm}
\textit{NEXTA for light curve simulation.}
\vspace{0.25cm}

NEXTA can also be applied to simulate occultation light curves, which can, for example, be used to evaluate data reduction software. The following is an example taken from the test video described in Sect. \ref{Sect. 2.4.1}. For this purpose, the state of the d\textsubscript{1s} LED (see Fig. \ref{Fig. 7} and Table \ref{Tbl. 1}) was photometrically measured with Tangra version 3.7.4. Figure \ref{Fig. 8} shows the light curve obtained in this way. We analysed the light curve with Occult/AOTA\footnote[1]{\url{http://www.lunar-occultations.com/iota/occult4.htm}} version 4.2022.5.12, see Fig. \ref{Fig. 9}. AOTA's analysis confirms the correct work of NEXTA, since D and R both occur at the full UTC second.
\subsubsection{NEXTA tests with a QHY174M-GPS camera}
\label{Sect. 2.3.2}
While the WATEC WAT-910HX-RC video camera used in the previous section offered a temporal resolution of only 20 ms, the QHY174M-GPS camera (Fig. \ref{Fig. 3}) achieves precise timings down to 1 $\mu$s. This camera is currently the only internally GNSS-controlled type and is often used for recording stellar occultations, also for professional campaigns \cite{buie_size_2020, strauss_sizes_2021}. The 2 Mega pixels 1920 x 1200 CMOS camera with a pixel size of 5.86 $\mu$m is equipped with a global shutter that enables exposure times from 900 s down to 5 $\mu$s.

However, exposure times in the range below about 0.5 ms are often not really necessary for occultation recordings, and the usual hardware chains do not allow them either due to the limited frame rates, especially because of the commonly used USB data lines. Primarily to test the faster time resolution digits of NEXTA, QHY174M-GPS FITS sequences with exposure times down to 0.1 ms were achieved by reducing the image size and using 8 bits of image resolution instead of the possible 16 bits. Despite these measures and also with a relatively fast PC (W7-64bit i7 16GB RAM), under 0.5 ms exposure time, the frame timing errors were close to 100\%, so that the fastest resolution of NEXTA of 0.1 ms could only partly be tested.

With SharpCap, due to a special calibration routine and the shutter control directly derived from the GNSS PPS signal, the QHY174M-GPS camera is able to determine $\mu$s-accurate time stamps for the start and end of each frame and write them to the corresponding FITS keyword. The time stamps of a single frame are thus determined exactly, even though frame failures may occur before or after due to insufficient USB connections or limited achievable frame rates.

\vspace{0.5cm}
\textit{Tests with short exposure FITS captures.}
\vspace{0.25cm}

The following example refers to a single FITS frame with an exposure time of 0.1 ms. To map the NEXTA display, the camera was equipped with a 35 mm lens 1:2.8. To capture we used SharpCap running  on a W7-64bit i7 16GB RAM PC. Figure \ref{Fig. 10} shows the test setup. Figure \ref{Fig. 11} presents the FITS recorded NEXTA display as well as the related time-relevant FITS keywords and demonstrates the agreement of the 0.1 ms resolved NEXTA readout with the temporal precision of the QHY174M GPS camera.

For further testing, a dropped frame free section of 431 frames was used from a 30 s FITS sequence with a frame exposure time of 1 ms. 
Figure \ref{Fig. 12} shows the first 4 and the last 3 images of the 431-image sequence. The NEXTA visual \mbox{timestamps} decoded from the images are compared to the corresponding FITS keywords provided by SharpCap.
From Figs. \ref{Fig. 11} and \ref{Fig. 12} can be concluded the camera's $\mu$s-precise timing which is confirmed by NEXTA. At the same time, this demonstrates NEXTA's suitability for carrying out such tests.

\vspace{0.5cm}
\textit{PyMovie/PyOTE and NEXTA.}
\vspace{0.25cm}

Besides Tangra/AOTA (Sect. \ref{Sect. 2.4.1}), the programs PyMovie and PyOTE have established themselves in the data reduction of occultation recordings. With PyMovie, the NEXTA's digits LED states can be read out (but not automatically decoded to analogue numbers) and written together with the camera's FITS header time stamps to a CSV file. Opening this file in PyOTE provides analysis options for both the occultation recording hardware and the associated data reduction software.

To demonstrate this, the QHY174M-GPS recorded FITS sequence (431 1 ms frames) described above was used. PyMovie was applied to photometrically record the brightness of the respective d-LEDs of the digits 1 s to 0.001 s. As can be seen from Fig. \ref{Fig. 13}, PyMovie/PyOTE are well suited tools to measure the LED states of NEXTA. The blue 0.1 s plot in Fig. \ref{Fig. 13} was considered as a simulation of an occultation drop and resolved with PyOTE; the result is shown in Fig.~\ref{Fig. 14}.

Figure \ref{Fig. 14} shows the very close match of the LED status with the GNSS time reference derived camera time stamps. Figure  \ref{Fig. 15} gives an analogous solution for the light curve of the 0.01 s digit (light blue in Fig. \ref{Fig. 13}). Its D time is identical to the D time of the 0.1 s light curve in Fig. \ref{Fig. 14}, confirming the precise work of NEXTA.
\subsubsection{NEXTA for the detection and measurement of rolling shutter effects}
\label{Sect. 2.3.3}
In contrast to the global shutter cameras described in previous sections, also rolling shutter cameras are used for recording stellar occultations. Modern CMOS cameras usually have electronic rolling shutters. Depending on the shooting parameters, rolling shutter cameras can cause image effects such as distortion of fast moving objects. The latter are not the main problem when recording stellar occultations, but in addition, with rolling shutters timing problems can occur due to the camera's sequential row-by-row sensor readout. The time data of an occultation derived from such recordings may therefore depend on the vertical sensor position of the occulted star.

With NEXTA, it is possible to determine if a camera has a rolling shutter, as this is not always immediately known. If a macro lens is used and LED images are sufficiently large it is also possible to use NEXTA to determine the readout rate of individual rows. This can be used, for example, to verify manufacturer's specification, test the frame-by-frame consistency of readout rate and determine exposure delay of a pixel row used for recording an occultation. It can also be used to convert the NEXTA optical time measured for a selected, individual pixel row to the first pixel row and compare is with image timing from FITS header, just like with global shutter (see Sect. \ref{Sect. 3.1}).

\vspace{1cm}
\textit{RunCam Night Eagle 3.}
\vspace{0.25cm}

Low-cost cameras are needed for mobile, unattended deployment of various occultation recording stations, for example in campaigns where the shadow path needs to be relatively densely populated and consequently a larger number of stations are required.

The RunCam Night Eagle 3\footnote[1]{\url{https://shop.runcam.com/runcam-night-eagle-3/}}, actually a first-person view (FPV) camera, meets this requirement and provides sufficient sensitivity comparable to the WAT-910HX-RC when its frame integration is not used \cite{ne3-report}. Depending on the RunCam Night Eagle 3 settings, this CMOS camera outputs a PAL or NTSC video signal. Additional timing equipment is required for recording occultations. Tests with NEXTA confirmed the presence of a rolling shutter on the camera (Fig. \ref{Fig. 16}).
The rolling shutter caused instrumental delays of the RunCam Night Eagle 3 were determined by \cite{ne3-report}. These delays can reach up to 16.7 ms for NTSC. PyOTE is able to incorporate the rolling shutter effects of this camera.

\vspace{2cm}
\textit{ZWO ASI1600MM camera.}
\vspace{0.25cm}

For the tests a ZWO ASI1600MM\footnote[2]{\url{https://astronomy-imaging-camera.com/product/asi1600mm-kit}} camera (16 Mega pixels 4656 x 3520 CMOS sensor, pixel size 3.8 $\mu$m) was equipped with a 1.8/50mm photo lens. Figure \ref{Fig. 17} demonstrates the rolling shutter effect. The image was taken with SharpCap with an exposure time of 50 $\mu$s. In the live view of SharpCap, vertically moving light patterns (related to Fig. \ref {Fig. 5}) within individual NEXTA LED sections indicate the presence of the rolling shutter.

Literature shows that the camera is used to record stellar occultations \cite{santos-sanz_physical_2022, vara-lubiano_multichord_2022}, although, unlike the QHY174M-GPS, additional timing equipment is required. It is not known if there is any effort or data on how to handle the camera's rolling shutter in the context of stellar occultations.

With NEXTA, however, an attempt was made to determine the magnitude of the readout delay over the entire sensor in the vertical direction. To realize this, the NEXTA display was mapped to the lower vertical end of the camera sensor. At the same time, the input of a fibre optic cable was placed in front of the d\textsubscript{0.1s} LED and the fibre optic cable's other end was positioned to be imaged on top of the sensor (Fig. \ref{Fig. 18}). Due to the relatively large sensor region of interest required (see Fig. \ref{Fig. 18}), the achievable frame rate was not greater than 46~FPS for bin2 and 23 FPS without binning. Therefore, only the NEXTA sections of 1 s and 0.1 s were time resolved and consequently only these sections could be used.

The results from the upper end of the sensor were found to be time delayed compared to the lower end. As shown in Fig.~\ref{Fig. 19}, the time difference was 1 frame (44 ms) at native sensor resolution and 22 ms correspondingly at x2 binning. As to expect, an analogous test with the QHY174M-GPS global shutter camera did not show a time delay.

We also tested the rolling shutter effect of the ZWO ASI1600MM camera using the methods described in Sect. \ref{Sect. 3}. As presented in Fig. \ref{Fig. 20} the readout time for a single row was measured to be 13.3 µs in bin1 mode. For the entire sensor follows a readout time of 46.8 ms. This result is in good agreement with the outcome of the measurement using a fibre optic cable (Figs. \ref{Fig. 18} and \ref{Fig. 19}). During the rolling shutter tests of the ZWO ASI1600MM camera, possible time delays were not in view because no external GPS device was available. 
\section{Tests of NEXTA on satellite tracking cameras}
\label{Sect. 3}
In this section we present results of image timing analysis with NEXTA for four different cameras, that are a potentially interesting choice for satellite tracking and survey observations. Two of them are using rolling shutter only, two other have software selectable shutter mode: rolling or global. All of them can be equipped with an external timing device for improved image timing accuracy.

\subsection{QHY 600M Pro}
\label{Sect. 3.1}
The QHY 600M Pro\footnote[1]{\url{https://www.qhyccd.com/scientific-cooled-camera-qhy600pro-imx455-cmos/}} camera (61 Mega pixels, 9600 x 6422 CMOS sensor, pixel size 3.76 $\mu$m) is one of a few astronomical cameras available on the market that can be supplied with a dedicated GNSS based timing device - GPSBOX. Its purpose is to provide accurate image timing with the resolution of 0.1$\mu$s and accuracy not specified clearly by the manufacturer. During the tests the camera was equipped with 1.4/50mm photo lens with macro extension rings. We used the fibre data interface therefore we collected data at higher FPS than possible using only USB3 interface (see Table \ref{Tbl. 2}). Exposure time was set to 70$\mu$s, the shortest possible for this camera, which caused slight blurring of the 100$\mu$s section of NEXTA. SharpCap software (ver. 4.0.9063 64-bit) and so called Live View mode, in which camera is continuously displaying images even if they are not commanded or recorded, was used throughout the tests. The alternative Still mode was not used, since it seems to reduce frame rates significantly.

QHY 600M Pro uses rolling shutter only, so the image timing provided by the camera (in FITS header) corresponds to its first row of pixels. All tests were performed using the central part of the sensors (see Fig. \ref{Fig. 21}), so direct comparison of image timings was not possible. Therefore we used NEXTA first to calculate the time delay between consecutive readouts of pixel rows (see Fig. \ref{Fig. 22}). Afterwards the obtained delay was used to convert the measured optical time for selected central row to the time of the first row. The time delay between readout of consecutive unbinned rows was 39.1$\mu$s in Photographic, High Gain and Extended Fullwell modes. It was doubled to 78.1$\mu$s in 2CMS mode and reduced to 15.6$\mu$s in 14 bit mode, which is only available when using optical fibre interface.

The time provided by QHY 600M Pro camera, equipped with a GPSBOX, for the beginning of the first pixel row was always slightly before the actual UTC time measured with NEXTA. The shift was between 0.5ms and 3.0ms, depending on camera mode (see Table \ref{Tbl. 2}). The time-bias in any particular mode was consistent throughout the tests. Although the deviation from the actual image timing is small, it is still an important correction that should be taken into account during satellite tracking observations of LEO targets.

\begin{table}[!h]
\caption{Timing analysis of QHY 600M Pro camera equipped with a GPSBOX.}         
\label{Tbl. 2}
\centering   
\begin{tabular}{l c c r r} 
\hline\hline
camera mode         & bin & fps & row readout & time-bias \\
                    &     &     & [$\mu$s]        & [ms] \\
\hline
Photographic        & 1x1 & 4   & 39.1        & -1.5 \\
Photographic        & 4x4 & 4   & 156.3       & -1.5 \\
High gain           & 1x1 & 4   & 39.1        & -1.5 \\
High gain           & 4x4 & 4   & 156.3       & -1.5 \\
Extended fullwell   & 1x1 & 4   & 39.1        & -1.5 \\
Extended fullwell   & 4x4 & 4   & 156.3       & -1.5 \\
Extended 2CMS       & 1x1 & 2   & 78.1        & -3.0 \\
Extended 2CMS       & 4x4 & 2   & 312.5       & -3.0 \\
14bit (fibre only)  & 1x1 & 8   & 15.6        & -0.5 \\
14bit (fibre only)  & 4x4 & 8   & 62.1        & -0.5 \\
\hline
\end{tabular}
\end{table}

Image timing of QHY 600M Pro was significantly worse and less consistent without the GPSBOX attached. The difference between the time recorded in FITS files and the actual time of exposure in this case was changing between 250ms and 750ms. Such results were achieved even with PC system clock synchronised using NTP (Network Time Protocol) over the internet with an accuracy of several milliseconds. It is a great example of unpredictable and in the case of satellite tracking unacceptable delays that are introduced even if a sensor is equipped with an electronic shutter.

\subsection{Andor Zyla 5.5}
\label{Sect. 3.2}
Andor Zyla 5.5\footnote[1]{\url{https://andor.oxinst.com/products/scmos-camera-series/zyla-5-5-scmos}} camera (5.5 Mega pixels, 2560 x 2160 CMOS sensor, pixel size 6.5 $\mu$m) is one of a few astronomical cameras available on the market that can operate in software selectable rolling and global shutter modes. It is equipped with an general purpose IO port. A trigger-out signal is generated at the beginning of each exposure which can be used to measure image timing independently on PC software. Unfortunately, Andor does not offer timing accessories similar to GPSBOX from QHY. Therefore an external GNSS image timing device for this camera (designed by KK) was used. During the test the camera was equipped with a 1.4/16mm photo lens. Exposure time was set to 27$\mu$s. Andor Solis\footnote[1]{\url{https://andor.oxinst.com/products/solis-software/}} software (ver. 4.32.30004.0), Single Scan mode and 16-bit dynamic range was used throughout the test. In rolling shutter mode the same procedure was used as described in Sect. \ref{Sect. 3.1}.

In rolling shutter mode the time delay between readout of consecutive unbinned rows of Zyla was 25.6$\mu$s at 200MHz and 9.1$\mu$s at 560MHz readout speed. This is very close to the manufacturer specification of 25.41$\mu$s and 9.24$\mu$s, respectively. The image timing of this camera recorded using external GNSS clock was always slightly behind the actual UTC time measured with NEXTA. The time-bias value was about 55ms at 200MHz and 20ms at 560MHz readout speed (see Table \ref{Tbl. 3}). This is significantly larger than in the case of QHY 600M Pro, but constant throughout the tests, therefore it is easy to apply corrections. In case of satellite tracking or survey these timing delays are necessary to be corrected for on all orbital regimes.

\begin{table}[!h]
\caption{Timing analysis of Andor Zyla 5.5 camera equipped with a GNSS timing device.}
\label{Tbl. 3}
\centering   
\begin{tabular}{l c c r} 
\hline\hline
camera mode         & bin  & row readout & time-bias \\
                    &      & [$\mu$s]    & [ms] \\
\hline
rolling 200MHz      & 1x1  & 25.6        & 55.4 \\
rolling 200MHz      & 2x2  & 50.0        & 54.3 \\
rolling 560MHz      & 1x1  & 9.1         & 19.8 \\
rolling 560MHz      & 2x2  & 18.2        & 19.8 \\
global 200MHz       & 1x1  & -           & <10 \\
global 560MHz       & 1x1  & -           & <10 \\
\hline
\end{tabular}
\end{table}

In global shutter mode we encountered unexpected difficulties. Even with a very short 27$\mu$s exposure time we have always seen the 0.1ms and 1ms sections of NEXTA with all LEDs lit on. The camera behaves as it would not fully block the incoming light during the readout of the sensor. According to specification the readout takes from 9.98ms to 27.44ms, depending on selected readout speed. See Fig. \ref{Fig. 23} for examples of NEXTA images in rolling and global shutter modes. The observed camera inability to take short exposures in global shutter mode reduced the resolution of the test to about 10ms. Therefore we were not able to test the camera's global shutter mode usability when the accuracy of image timing is required to be better than 10ms.

Image timing of Andor Zyla 5.5, without an external GNSS clock, is saved by Andor Solis software with resolution of only 1 second in FITS headers. The difference between the time recorded and the actual time of the exposure in this case was usually within 1 second, as expected. Surprisingly, examples of much larger differences, up to 4 seconds, were also encountered. This renders the software image timing practically useless for any of the applications discussed in this paper. It is worth noting that using our own software, based on Linux SDK for Andor cameras, we were able to significantly improve the software image timing accuracy, down to the level of a few milliseconds.

\subsection{Andor Balor}
\label{Sect. 3.3}
Andor Balor 17-12\footnote[1]{\url{https://andor.oxinst.com/products/scmos-camera-series/balor-scmos}} (16.9 Mega pixels, 4128 x 4104 CMOS sensor, pixel size 12 $\mu$m) is a large format, high FPS camera (up to 54Hz full frame) capable to operate in software selectable rolling and global shutter modes. It is equipped with a dedicated IRIG-B port for connecting a compatible GNSS receiver, however we did not use it. Instead we used the trigger-out functionality of the general purpose IO port just as with Andor Zyla (see Sect. \ref{Sect. 3.2} ). Tests were performed using Andor Solis software and 0.11ms exposure time (Fig. \ref{Fig. 24}).

The time-bias measured with NEXTA is presented in Table \ref{Tbl. 4}. In global shutter mode it was below the 0.1ms resolution of NEXTA - perfect result for even the most demanding satellite observations. In rolling shutter mode, however, we detected that the GNSS timing for the first row of pixels was always recorded 1.5ms prior to the actual beginning of image exposure. This time-bias was constant and therefore easily reducible. As in the case of Andor Zyla, the timing provided by Andor Solis software for Balor was only with the resolution of 1 sec.

We did not observed the same problem of "light leaking" through closed electronic shutter in global shutter mode as in Andor Zyla 5.5. This shows that the problem was most likely not related to the procedure or equipment used during the test but a camera itself.

\begin{table}[!h]
\caption{Timing analysis of Andor Balor 17-12 camera equipped with a GNSS timing device connected to IO port.}
\label{Tbl. 4}
\centering   
\begin{tabular}{l c c r} 
\hline\hline
camera mode         & bin  & row readout & time-bias \\
                    &      & [$\mu$s]    & [ms] \\
\hline
rolling             & 1x1  & 5.49        & -1.5 \\
global              & 1x1  & -           & 0.0 \\
\hline
\end{tabular}
\end{table}

\subsection{FLI Kepler}
\label{Sect. 3.4}

FLI Kepler 4040\footnote[2]{\url{https://www.flicamera.com/kepler/kepler.html}} (16.9 Mega pixels, 4096 x 4096 CMOS sensor, pixel size 9 $\mu$m) is a large format, high FPS camera (up to 20Hz full frame) with a popular, front-illuminated GSense4040 sensor. The camera has an electronic rolling shutter and is equipped with a general purpose IO port with trigger-out functionality. We did not have the FLI Kepler Image Time Stamp device, which is claimed to provide image timing accuracy of 1.5ms. Instead we used our own timing device which has similar accuracy and uses the same signals from the camera for measurements. The exposure time used here was 41$\mu$s, software was a custom CLI solution based on FLI SDK\footnote[1]{\url{https://www.flicamera.com/software/index.html}}.

Only one camera mode (High Dynamic Range + Low Dark Current), with 1x1 binning was tested with NEXTA. The single row readout time was measured as 10.4$\mu$s, and time-bias for the first row of pixels with respect to image timing based on GNSS receiver was consistently 0.3ms. When compared to software based image timing the time-bias was varying between about 40ms and 90ms from image to image.

\begin{figure}
   \centering
   \includegraphics[width=10cm]{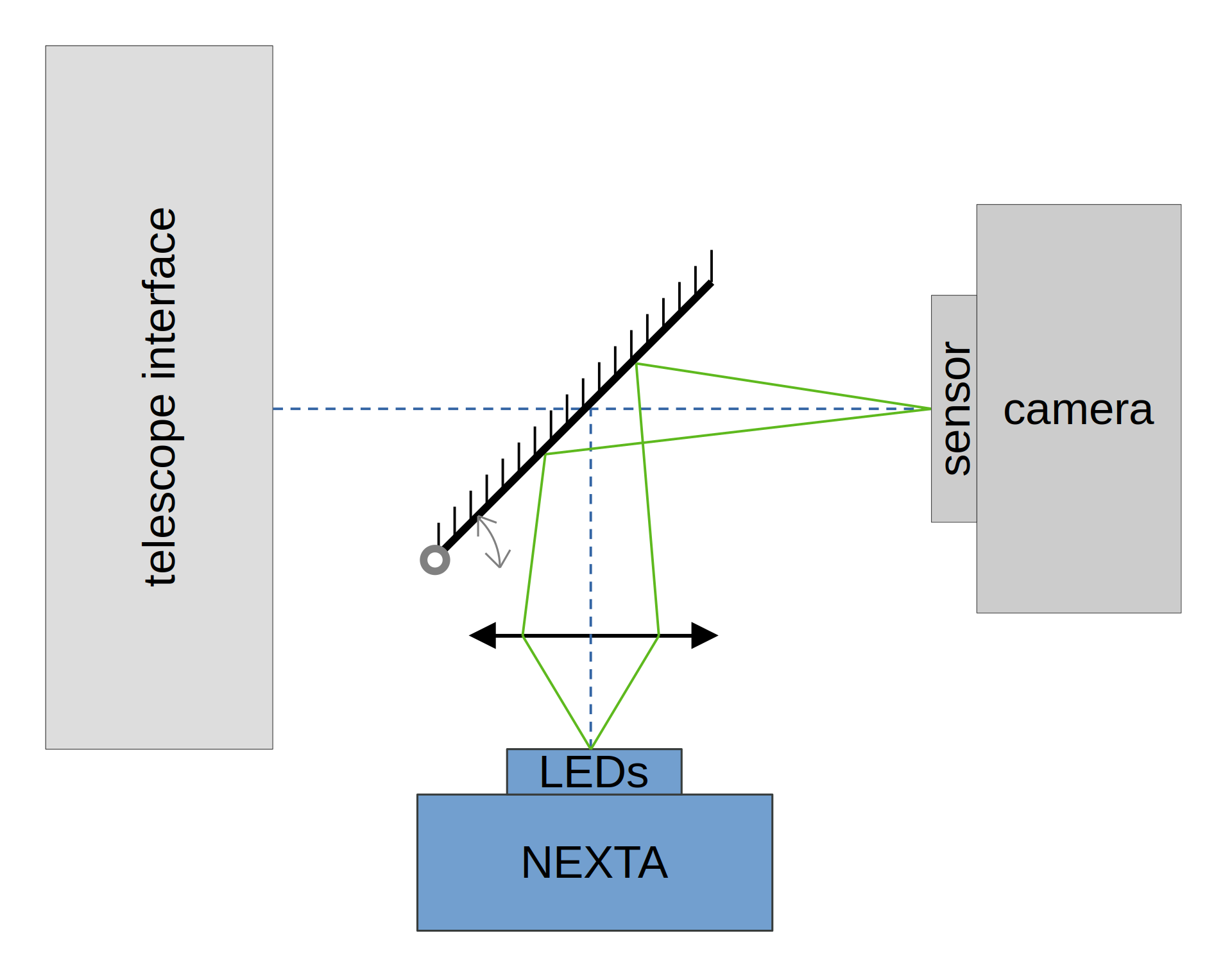}
   \caption{Schematic illustration of an idea to use NEXTA directly between the telescope and camera. A folding mirror and a focusing lens would be necessary, so a sufficiently large back-focus distance is required. Dashed lines represent optical axis of the telescope and the lens, green lines show selected light rays. The folding mirror is presented in open position, during measurements with NEXTA. During observations it would be in a closed position which does not obstruct light coming from the telescope. This idea should allow for a very convenient use of NEXTA for example during telescope slewing.}
   \label{Fig. 25}
\end{figure}

\section{Conclusions}
\label{Sect. 4}

NEXTA has proven to be a very suitable tool to test the timing accuracy of various image timing systems and, to some extent, the associated data reduction software. One of the advantages of the instrument is its simplicity and ease of reproducibility. In a large number of tests with a wide range of devices, the NEXTA showed no problems and provided a valuable insight into the image timing precision and accuracy.

The primary limitation of the NEXTA is that the resolution of the measurement possible with this device is limited to the minimum exposure time of the camera which is being tested. Therefore it is very well suited for high FPS devices equipped with an electronic shutter and less so for low FPS sensors equipped with mechanical shutters.

NEXTA allows to significantly improve the accuracy of determination of time-bias when compared to a classic method utilizing observations of navigation satellites. The former has two orders of magnitude better resolution and accuracy than the latter and allows to make calibration measurements during the day. It is possible to install the NEXTA, after some adaptations, directly on the telescope and use it during telescope slewing (see Fig. \ref{Fig. 25}). It is also possible to dedicate part of a large sensor field of view for NEXTA permanently, but that would require some adaptation in order to minimise the risk of overexposure for longer exposure time. Both solutions would allow for a much more frequent calibration measurements and therefore better monitoring of camera image timing system accuracy and stability.

The camera tests that we conducted comprised mostly of units equipped with CMOS sensors and only electronic shutters. They showed that software based image timing is accurate at best at the level of tens of ms and sometimes only at the level of full seconds. Thanks to NEXTA, we were able to prove that only the internally directly PPS-controlled QHY174M-GPS camera meets its µs-accurate specification. When using external GNSS based image timers attached to trigger-out port, we see significant improvements in image timing accuracy. Nevertheless, only in the case of Andor Balor working in global shutter mode we observed correct image timing provided by the external GNSS timer. All other cameras (including Balor working in rolling shutter mode) had measurable time-biases. They range from 50 ms to -3 ms and were stable during the short term of the conducted tests. These are non negligible corrections that should be taken into account when measuring LEO satellites and, to a lesser degree, also asteroid occultations.

The open question that was not tested is a long term stability of image timing systems. With the NEXTA being cheap and easy to manufacture, and accurate to the level of 0.1 ms, monitoring of such stability should become much more widely available.

The authors thank B. Anderson for the development of \href{https://pypi.org/user/BobAnderson52/}{PyMovie/PyOTE}, H. Pavlov for the \href{http://www.hristopavlov.net/Tangra/Tangra.html}{Tangra} development, B. Herald, developer of \href{http://www.lunar-occultations.com/iota/occult4.htm}{Occult/AOTA} and R. Glover at \href{https://www.sharpcap.co.uk/}{AstroSharp Limited}, developer of SharpCap.
This work was supported by the National Science Centre, Poland, through grant no. 2020/39/O/ST9/00713

\section*{References}

\bibliographystyle{iopart-num}
\bibliography{bibliography.bib}

\begin{appendix}                             
\section{Technical details on NEXTA construction and operation}
\label{App. A}

\begin{figure}[hbt!]
   \centering
   \includegraphics[width=14cm]{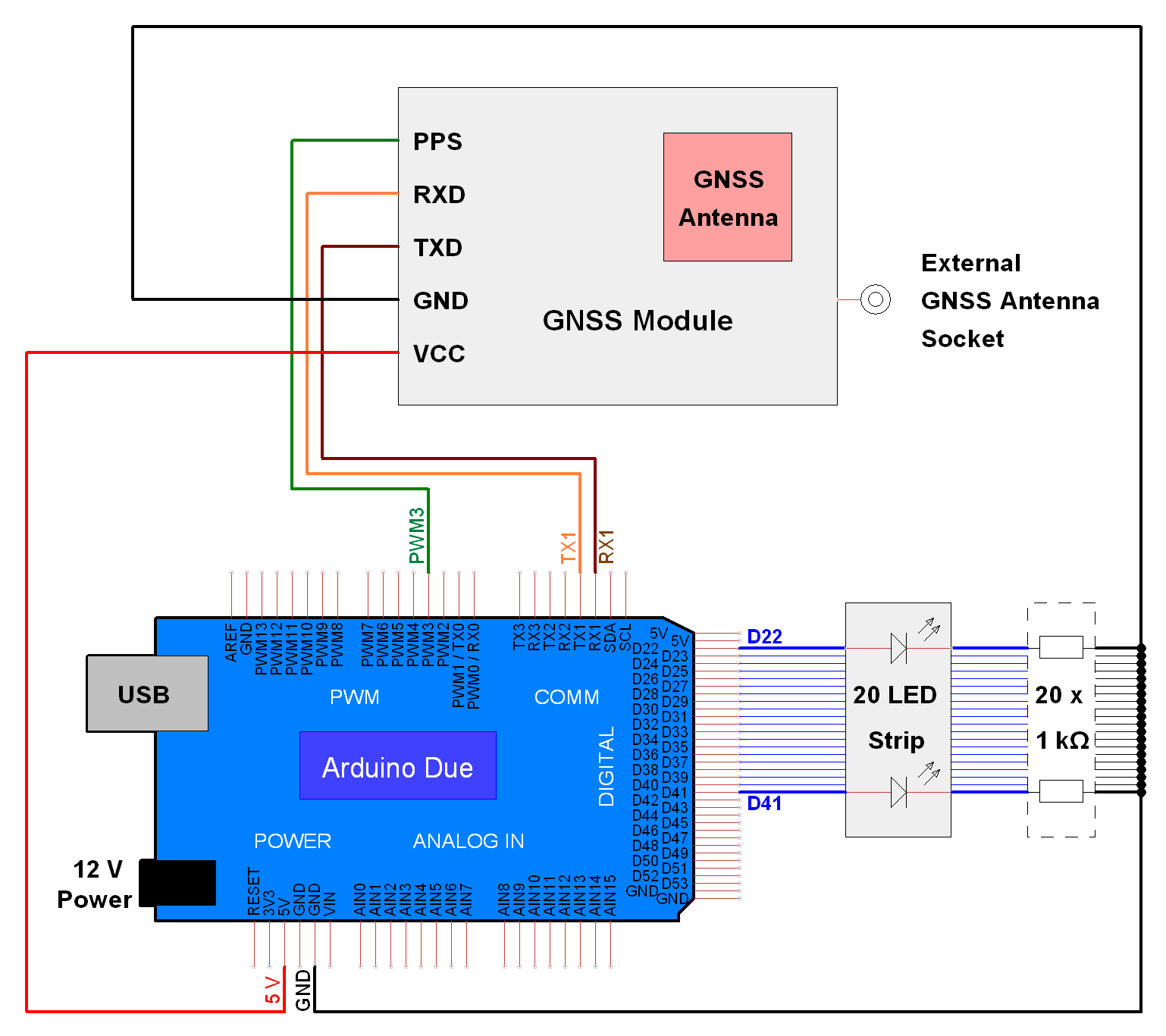}
   \caption{NEXTA wiring diagram. When using a GNSS module with 3.3~V power supply, this must be connected to the Arduino Due 3V3 pin.}
   \label{Fig. 26}
\end{figure}

Figure \ref{Fig. 26} presents the NEXTA wiring diagram. Figures \ref{Fig. 1} and \ref{Fig. 2} show a parts overview and a NEXTA layout example. Besides the Arduino Due controller board, the second main component is a GNSS receiver that provides the PPS (one pulse per seconds) signal. This output is reconfigured to PP10S (one pulse per 10 seconds) during the operation of the device. There are many variants of these receivers on the market, but it is important to select one equipped with battery and compatible with UBLOX chip of at least the NEO-7M version. The battery is necessary so that information about leap seconds is stored inside and UTC time displayed correctly.

There are no special requirements for the other components. We recommend the use of an Arduino Proto Shield (the one from Arduino Mega is also compatible) to avoid soldering on the actual Arduino Due board. Furthermore, it makes sense to carry out a test breadboard assembly before the actual assembly.

Basic check of validity of impulses coming from a GNSS module was implemented into the firmware code of NEXTA. Nevertheless it is still possible that an electro-magnetic interference (EMI) would trigger the device to display erroneous time for up to 10s. Therefore some care should be taken in order to lower the risk of such a situation. For example the device should be completely static (GPS wire movements can trigger false impulse) and no unnecessary electric devices should be operated nearby.

\section{NEXTA firmware}
The firmware code for NEXTA is available here \url{http://www.astro.amu.edu.pl/files/NEXTA.ino}. It assumes that the main board of NEXTA is Arduino Due and all connections are made in accordance with scheme presented in Fig. \ref{Fig. 26}. It was created and compiled in Arduino IDE 1.8.19. The Arduino Due is not supported by default and requires user to download Arduino SAM Boards (we used version of 1.6.12) using Boards Manager.
\label{App. B}

\section{Information and error codes}
\label{App. C}

\begin{figure}[hbt!]
   \centering
   \includegraphics[width=12cm]{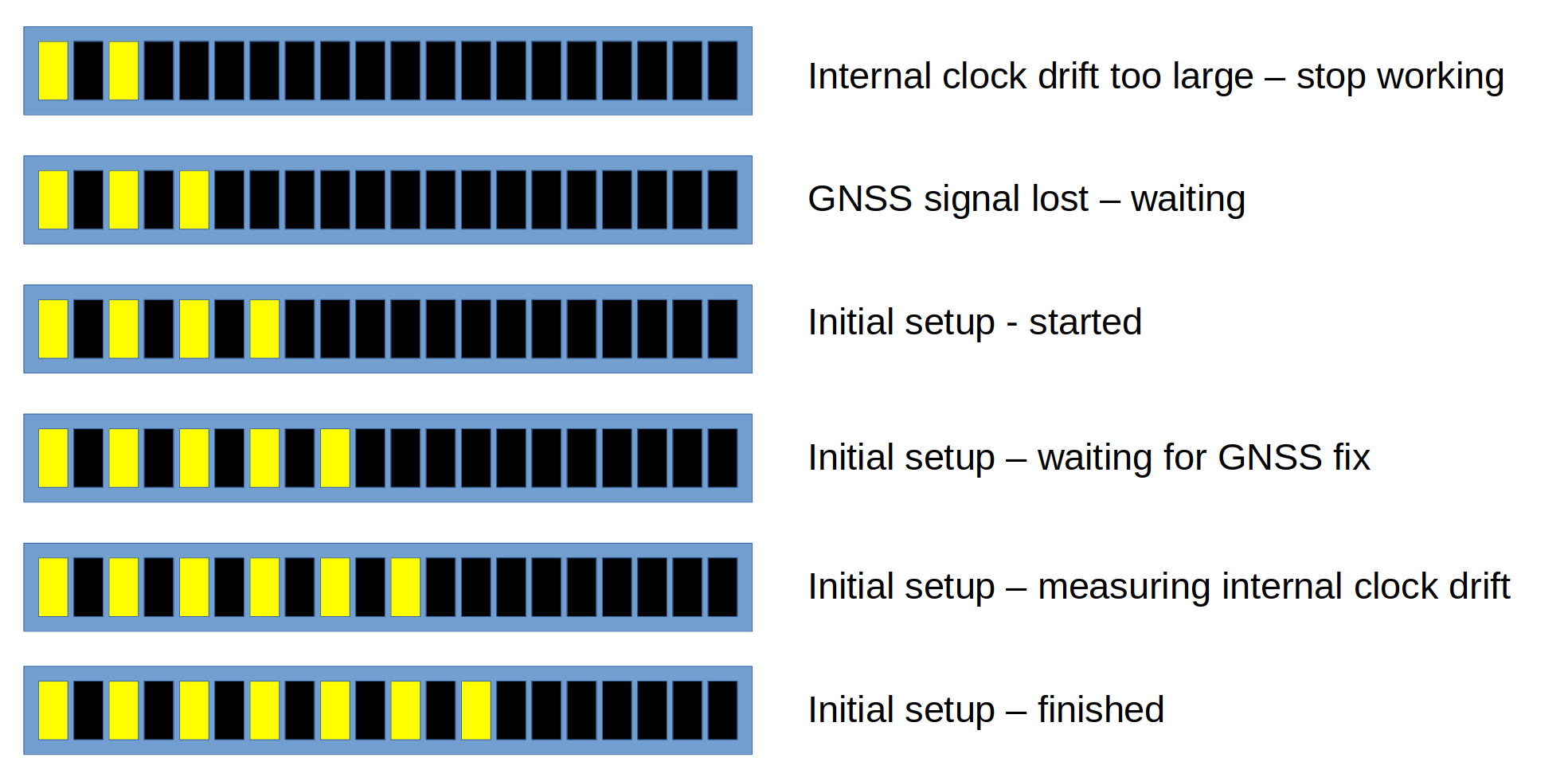}
   \caption{NEXTA information and error codes.}
   \label{Fig. 27}
\end{figure}

NEXTA can display several codes by turning on LEDs combinations which are impossible during the normal operation of the device. Most of them display information about the stage at which the initial setup is. They are presented in the Fig. \ref{Fig. 27}.

\end{appendix}

\end{document}